\documentclass[journal,twocolumn]{IEEEtran}

\usepackage{amsmath,amsfonts,amssymb,paralist,latexsym,mathtools,xparse,amsthm, epsfig,color,rotating,times,float,subcaption,algorithm,verbatim,enumitem,bm,graphicx,listings,url,multirow,amstext,array}
\usepackage{xifthen,ifthen}%
\usepackage{tikz} \usetikzlibrary{patterns,hobby,fit}
\usepackage{pgfplots}

\newcommand{\thdim}{D}

\newcommand{\obs}{y}
\newcommand{\obsl}{y_{l}}

\newcommand{\ep}{\epsilon}
\newcommand{\thl}{\theta_{l}}
\newcommand{\errl}{v_{l}}
\newcommand{\target}{l}
\newcommand{\numtarget}{L}
\renewcommand{\th}{\theta}

\newcommand{\z}{z}

\newcommand{\reg}{\psi}
\newcommand{\err}{v}

\newcommand{\state}{x}

\newcommand{\statedim}{X}

\newcommand{\tp}{P} 
 
\newcommand{\statespace}{\mathcal{X}}
 
\newcommand{\horizon}{N}

\newcommand{\Q}{Q}
\newcommand{\argmax}{\operatornamewithlimits{arg\,max}}
\newcommand{\argmin}{\operatornamewithlimits{arg\,min}}
\newcommand{\E} {\Bbb{E}}

\makeatletter
\newcommand{\al}{
  \alpha
  \@ifnextchar\bgroup{\sb}{}
}
\newcommand{\be}{
  \beta
  \@ifnextchar\bgroup{\sb}{}
}
\newcommand{\ga}{
  \gamma
  \@ifnextchar\bgroup{\sb}{}
}
\makeatother

\newtheorem{theorem}            {Theorem}
\newtheorem{lemma}            {Lemma}
\newtheorem{result}            {Result}

\newtheorem{definition}            {Definition}

\newcommand{\fun}{\phi}
\newcommand{\reals}{\mathbb{R}}

\NewDocumentCommand{\terminal}{e{_}ge{_}}{%
   {\tau}%
   \IfValueT{#2}{_{#2}}%
 }
 
\NewDocumentCommand{\terminalc}{e{_}ge{_}}{%
   {\tau}%
   \IfValueT{#2}{_{#2}}%
 }

\NewDocumentCommand{\da}{e{_}ge{_}}{%
   {d}%
   \IfValueT{#2}{_{#2}}%
 }

\newcommand{\diag}{\operatorname{diag}}

\newcommand{\belief}{\pi}

\newcommand{\p}{\prime}
\newcommand{\ones}{\mathbf{1}}

\newcommand{\filter}{T}
\newcommand{\filterd}{\sigma}

\newcommand{\oprob}{B}
\def \defn {\stackrel{\text{defn}}{=}}

\def\g1{\geq_{r_1}}

\newcommand{\pdf}{p}

\newcommand{\tht}{\theta^o}

\newcommand{\thlt}{\theta_l^*}

\newcommand{\sym}{S}

\newcommand{\Sig}{\Sigma}
\newcommand{\cov}{\operatorname{Cov}}
\newcommand{\var}{\operatorname{Var}}
\newcommand{\normal}{\mathbf{N}}

\newcommand{\lam}{\lambda}
\newcommand{\lamt}{\lambda^o}

\newcommand{\co}{\sym}

\newcommand{\M}{M_\target}
 \newcommand{\numtargetset}{[\numtarget]}

\newcommand{\regv}{\psi}
\newcommand{\I}{\mathcal{I}}
\newcommand{\nsym}{\bar{\sym}}
\newcommand{\bz}{\bar{z}}
\newcommand{\bsym}{\nsym}
\newcommand{\bth}{\bar{\th}}
\newcommand{\obsv}{\mathbf{\obs}}
\newcommand{\perm}{\sigma}
\renewcommand{\Re}{\mathfrak{Re}}
\newcommand{\mle}{\hat{\th}}
\newcommand{\Th}{\Theta}
\newcommand{\prior}{\pi_0}

\renewcommand{\colon}{\mathord{:}}
\newcommand{\bobs}{\bar{\obs}}

\newcommand{\conv}{\otimes}
\newcolumntype{L}{>{$}l<{$}} 
\newcommand{\sig}{\sigma}
\newcommand{\sigblank}{}
\newcommand{\Anon}{\perm}

\newcommand{\am}{\mathcal{A}}

\newcommand{\boprob}{\bar{\oprob}}
\newcommand{\kalmancov}{\Sigma}
\newcommand{\Pe}{P_e}
\newcommand{\blackwell}{\geq_{\mathcal{B}}}

\newcommand{\levels}{q}
\newcommand{\kerM}{M}
\newcommand{\Pavg}{P_e}
\newcommand{\dist}{\xrightarrow[]{D}}
\newcommand{\Reg}{\Psi}
\newcommand{\bSig}{\bar{\Sig}}
\newcommand{\R}{R}

\newcommand{\Bth}{\boldsymbol{\th}}
\newcommand{\Tr}{\operatorname{Tr}}
\newcommand{\citep}{\cite}
\newcommand{\citet}{\cite}

\newcommand{\mcstep}{\mu}
\newcommand{\hyper}{a}
\newcommand{\inty}{\int_{\mathcal{Y}}}
\newcommand{\intyb}{\int_{\mathcal{\bar{Y}}}}

\tikzset{
    block/.style={rectangle, draw, line width=0.5mm, black, text width=4em, text centered,
                 minimum height=2em},
               line/.style={draw, -latex}}
\tikzset{
    blockb/.style={rectangle, draw, line width=0.5mm, black, text width=6.5em, text centered,
                 minimum height=2em},
               line/.style={draw, -latex}}
\tikzset{
    block2/.style={rectangle, draw, line width=0.2mm, black, text width=1em, text centered,
                 minimum height=10em},
    line/.style={draw, -latex}}     

\usepackage[labeled,resetlabels]{multibib}
\newcites{A}{References in Supplementary Document}

\allowdisplaybreaks
\title{Adaptive Filtering Algorithms for Set-Valued Observations--Symmetric Measurement Approach to Unlabeled and Anonymized Data} 
\author{Vikram Krishnamurthy,  {\em Fellow IEEE}, 
\today
  \thanks{Vikram Krishnamurthy is with the School of Electrical and Computer Engineering, Cornell University, Ithaca, NY, 14853, USA.
    Email: vikramk@cornell.edu.  This research was supported by the National Science Foundation grant CCF-2112457, U.S.\ Army Research Office  grant
    W911NF-21-1-0093 and US. Air Force Office of Scientific Research grant FA9550-22-1-0016.}}

\begin{document}

\maketitle

\begin{abstract}
  Suppose $\numtarget$ simultaneous independent stochastic systems generate observations, where the observations from each system depend on the  underlying parameter of that system. The observations are unlabeled (anonymized), in the sense that an analyst  does not know which observation came from which stochastic system. How can the analyst estimate the underlying parameters of  the $\numtarget$ systems?
Since   the anonymized observations at each time  are an unordered set of $\numtarget$ measurements (rather than a vector), classical stochastic gradient algorithms cannot be directly used.  
By using  symmetric polynomials, we formulate  a symmetric measurement equation   that maps  the  observation set  to a unique vector. By exploiting that fact that the algebraic ring of multi-variable polynomials is a unique factorization domain over the ring of one-variable polynomials,    we construct  an adaptive filtering  algorithm that yields a  statistically consistent estimate of the
underlying parameters.
We  analyze the asymptotic covariance of these estimates to quantify the effect of anonymization.
Finally, we characterize the  anonymity of the observations in terms of the  error probability of the maximum aposteriori  Bayesian estimator. Using  Blackwell dominance of mean preserving spreads, we  construct  a partial ordering of the noise densities which
relates  the anonymity of the observations to the asymptotic covariance of the  adaptive filtering algorithm.
\end{abstract}


\section{Introduction} \label{sec:intro}

The classical stochastic gradient algorithm operates on a {\em vector-valued} observation process that is inputted to the algorithm at each time instant.  Suppose due to anonymization,  the observation  at each time is  a {\em set} (i.e., the elements   are unordered  rather than a vector). Given these anonymized observation sets over time, how to construct a stochastic gradient algorithm to estimate the underlying parameter?

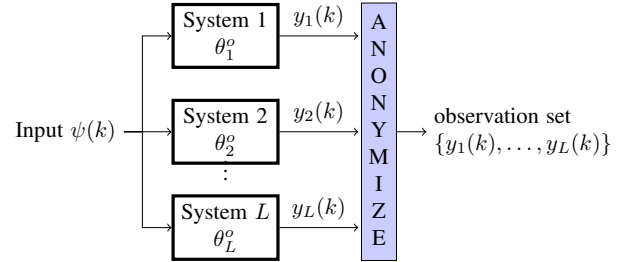
\begin{figure}[h]
  \centering
\scalebox{0.85}{\begin{tikzpicture}[node distance = 1cm, auto]
    \node [block] (S1) {System~1 $\tht_1$};
    \node [block, below of=S1,node distance=1.5cm] (S2) {System~2 $\tht_2$};
   \node[right of = S1, node distance=2.2cm] (anon1) {};
   \node[right of = S2, node distance=2.2cm] (anon2) {};

  \node [below of=S2,node distance=0.5cm] (vdots) {$\vdots$};
   
     \node [block, below of=S2,node distance=1.5cm] (SL) {System~$\numtarget$ $\tht_\numtarget$};
   \node[right of = SL, node distance=2.2cm] (anon3) {};
    \node  [left of=S2,node distance=2.5cm] (Input) {Input $\reg(k)$};

    \node[right of = S2, node distance=2.4cm, draw,fill=blue!20] (anon) {
\begin{tabular}{@{}c@{}}
A\\
N\\ O \\ N \\ Y \\  M \\ I \\ Z \\ E  
\end{tabular}
};
\node [right of=anon,text width=15em,node distance=3.5cm] (aobs) {observation set \\ $\{y_1(k), \ldots,y_\numtarget(k)\}$};

\draw[->] (anon) -- (aobs);

\draw[->] (S1) -- node[above]{$\obs_1(k)$}(anon1);
\draw[->] (S2) -- node[above]{$\obs_2(k)$}(anon2);
\draw[->] (SL) -- node[above]{$\obs_\numtarget(k)$}(anon3);

\draw[->] (Input) -- ++ (1.2,0) |- (S1.west);
\draw[->] (Input) -- ++ (1.2,0) |- (SL.west);
\draw[->] (Input) -- ++ (1.2,0) |- (S2.west);

\end{tikzpicture}}
\caption{Schematic setup comprising $\numtarget$ stochastic systems. Given the sequence of anonymized observation sets $(\{\obs_1(k),\ldots,\obs_\numtarget(k)\}, k=1,2,\ldots)$, the aim is to estimate the underlying parameter set $\tht = \{\tht_1,\ldots\tht_\numtarget\}$ of the $\numtarget$ systems.}
\label{fig:anon}
\end{figure}


Figure~\ref{fig:anon} shows the schematic setup comprising $\numtarget$ simultaneous independent  stochastic systems
indexed by $\target=1,\ldots, \numtarget$, evolving over discrete time $k=1,2,\ldots$. Each stochastic system $\target$  is  parametrized by true model $\tht_l \in \reals^\thdim$ and  generates
observations $\obsl(k) \in \reals^\thdim$ given input signal $\thdim \times \thdim$ dimensional matrix $\reg(k)$:
\begin{equation}
  \label{eq:obs}
  \obsl(k) = {\reg(k)}\, \tht_\target + \errl(k), \quad  \target \in \numtargetset \defn\{1,\ldots,\numtarget\}
\end{equation}
We assume that $\errl(k) \in \reals^\thdim$ is iid random sequence  with bounded second moment. 
We (the analyst)  know (or can choose) the input signal sequence $(\reg(k), k=1,2,\ldots)$. For convenience, assume that elements of 
$(\reg(k), k=1,2,\ldots)$ are zero mean iid  sequences of random variables.
Thus the output of the $\numtarget$ stochastic systems at time $k$ is the observation {\bf matrix} 
$$\obsv(k) = [\obs_1(k),\ldots,\obs_\numtarget(k)]^\p \in \reals^{\numtarget\times \thdim}$$
where  $a^\p$  denotes transpose of matrix $a$.

The analyst  observes at each time $k$ the anonymized (unlabeled) observation \textbf{set}
\begin{equation}
  \label{eq:setobs}
  \obs(k)=\Anon_k(\obsv(k)) = \{\obs_1(k),\ldots,\obs_\numtarget(k)\}
\end{equation}
The anonymization map $\Anon_k$ is a  permutation over the set $\{1,2,\ldots,\numtarget\}$.
By anonymization\footnote{For now  we use anonymization to denote masking the index label~$\target $ of the stochastic process.  Sec.~\ref{sec:applications} motivates this in terms of $k$-anonymity.} we mean that by transforming the matrix  $\obsv$ with ordered rows to set $\obs$ with unordered rows, the index label $\target \in \{1,2,\ldots,\numtarget\}$ is hidden; that is, the observations are unlabeled.
The time dependence of  $\Anon_k$ emphasizes that the permutation map operating on $\obsv(k)$ changes at each time~$k$.

{\em Aim}. The analyst only sees the anonymized observation set $\obs(k)$  at each time~$k$.
Given the time sequence of observation  sets $(\obs(k), k =1,2,\ldots)$, the aim of the analyst is to estimate 
the underlying set of true  parameters  $\tht = \{\tht_1, \ldots,\tht_\numtarget\}$ of the $\numtarget$ stochastic systems.  Note that the analyst aim is  estimate  the {\em set}  $\tht$; due to the anonymization (unknown permutation map), in general,  it is impossible to estimate which  parameter belongs to which stochastic system.

{\em Remarks}: (i) Another  way of viewing the  estimation objective is:  Given noisy measurements of unknown permutations of the rows a matrix, how to estimate the elements of the matrix? Our main result  is to propose a  symmetric transform
framework that circumvents
modeling the permutations $\perm_k$ and is completely agnostic to the probabilistic structure of $\perm_k$.

(ii) The assumption that $\reg(k)$ is a $\thdim\times \thdim$ matrix in~\eqref{eq:obs} is without loss of generality. The classical LMS framework involves scalar valued observations
$o_\target(k) = \regv^\p(k) \tht_\target + e_l(k)$ where 
$\regv(k) \in \reals^{\thdim}$ is the known regression vector, and $e_\target(k)$ is a noise process. If we stack $\thdim$ such scalar observations into
the vector $\obs_l(k)$, then we obtain~\eqref{eq:obs}.  

(iii) The model reflects \textit{uncertainty associated with the origin of the measurements} (arbitrary permutation)  in addition to their inaccuracy (additive noise).
If we knew which observation $m$ was associated with which stochastic system $l$, then
we can estimate  each $\thlt$ independently  as the solution of the following stochastic optimization problem:
$\th^*_\target = \argmin_\th \E\{ (\obsl(k) - {\reg(k)} \thl)^2\}$.
Then the classical  LMS algorithm can be applied to estimate each $\thlt$ recursively as:
\begin{equation}
  \label{eq:LMSclassical}
\thl(k+1) = \thl(k) + \ep\, \reg(k) \big( \obsl(k) - {\reg(k)}\, \thl(k) \big)
\end{equation}
where the fixed step size  $\ep > 0$ is a small positive constant.

(iv)
Since the ordering of the elements of the set $\{\obs_1(k),\ldots,\obs_\numtarget(k)\}$ is arbitrary,
 we cannot use the LMS algorithm~\eqref{eq:LMSclassical}.  If   we naively choose  a random permutation of the set  $\obs(k)$ as the observation vector, and feed  this $\numtarget$-dimensional observation vector into $\numtarget$ LMS algorithms~\eqref{eq:LMSclassical}, then the estimates will not in general converge to $\tht_\target$, $\target=1,\ldots,\numtarget$.

 (v) Finally, the above formulation only makes sense in the stochastic case. The deterministic case is trivial. If  the noise $\err_\target(k) = 0$ and  input matrix $\reg(k)$ is invertible, then we need only one observation $\obs$ to completely determine the parameter set $\tht$, regardless of the permutation $\perm_k$. 


 \subsection*{Stochastic Optimization with Anonymized Observations. Circumventing Data Association}

 Broadly, there are two classes of methods for dealing with unlabeled observation model~\eqref{eq:obs}, \eqref{eq:setobs}.
 One class of methods is based on data association~\cite{BLK08,MMW05,SFC90}.
%
 Data association  deals with the question:  How can the observations   from multiple simultaneous processes  be assigned to specific processes  when there is uncertainty about  which observation
came from which  process?
 Since the observations are anonymized wrt to the index label $\target$ of the random processes,
 one approach is to construct a classifier  that assigns at each time $k$ the observation $\obs_\target(k)$ to a specific process  $m$. Because the number of process/observation pairs grows combinatorially with  the number
of processes and observations, a brute force approach to the data association problem
is computationally prohibitive. 
Data association is  studied extensively  in  Bayesian filtering for target tracking.
 In this paper we are dealing with stochastic optimization instead of Bayesian estimation, where we wish to preserve the convex structure of the problem.  

The  second class of methods  bypasses data association, i.e.,  labels are no longer estimated (assigned) to the anonymized observations. This  paper focuses on using symmetric transforms to bypass data association, as discussed next.

\subsection{Main Idea. Symmetric Transforms \& Adaptive Filtering}

Since the assignment step in data association can destroy the convexity structure  of a stochastic optimization problem, a natural question is:    {\em Can data association be circumvented in a stochastic optimization problem?}
A remarkable approach developed in the 1990s by Kamen and coworkers \citep{Kam92,KS93}  in the context of Bayesian estimation, involves using  symmetric transforms. This ingenious idea  circumvents data association; see also~\cite{HBW17} and references therein.
In this paper we extend this idea of symmetric transforms  to  stochastic optimization. Specifically, we show that the symmetric transform approach preserves convexity. Since \citet{Kam92} deals with Bayesian filtering for estimating the state, convexity is irrelevant. In comparison,   preservation of convexity is crucial in  stochastic optimization problems to ensure that the  estimates of a stochastic gradient algorithm converge to the global minimum.

To explain our main ideas, suppose there are $\numtarget=3$ scalar-valued random processes, so each observation $\obs_\target(k)$ is scalar-valued. Further for simplicity assume the input signal $\reg(k) = 1$; so the observations are $\obs_\target(k) = \tht_\target + \err_l(k)$.
Given  the anonymized  observation set $\obs(k)=\{\obs_1(k),\ldots,\obs_3(k)\}$ at each time $k$, how to estimate the parameters $\tht_1, \tht_2, \tht_3$?  Our main idea is to  use the {\em set} $\obs(k)$ to construct a pseudo-measurement {\em vector} $\z(k) \in \reals^3$. Suppressing the time dependency $(k)$ for notational convenience, we  construct the  pseudo-measurements
 $z_1, z_2,z_3$ via a symmetric transform as follows:
 \begin{equation}
  \label{eq:sym3}
\begin{split}
    z_1 &= \sym_1\{\obs_1,\obs_2,\obs_3\}= \obs_1 + \obs_2 + \obs_3 \\
    z_2 &= \sym_2\{\obs_1,\obs_2,\obs_3\} = \obs_1 \,\obs_2 + \obs_1 \,\obs_3 +      \obs_2\, \obs_3 \\
  z_3 &=   \sym_3\{\obs_1,\obs_2,\obs_3\}  = \obs_1\,\obs_2\,\obs_3
\end{split}
\end{equation}
The key point is that the pseudo-observations $z_\target$ are symmetric in $\obs_1,\obs_2,\obs_3$.
Any permutation of the elements of $\{\obs_1,\ldots,\obs_3\}$  does not affect~$z_l$.
In this way, we have circumvented the data association problem; there is no need to assign (classify) an observation to a specific process. But we have introduced a new problem: estimating $\tht$ using the pseudo-observations is no longer a  convex stochastic  optimization problem.
To estimate $\tht$ we minimize the second order moments to compute:
\begin{align}
  \label{eq:objintro}
\th^*&= \argmin_\th\{ \E\{ (z_1 - (\th_1 + \th_2 + \th_3))^2\}  \\ &+\E\{ (z_2 - (\th_1 \th_2 + \th_1 \th_3 + \th_2 \th_3))^2\}
  + \E\{(z_3 - \th_1 \th_2\th_3)^2\} \} \nonumber 
\end{align}
Clearly the multi-linear objective~\eqref{eq:objintro}  is  non-convex in $\th_1,\th_2,\th_3$. However, the problem is convex in the symmetric transformed variables (denoted as $\lam$ below), and the original variables $\th$ can be evaluated by inverting the symmetric transform. We formalize this as follows:

\begin{result} (Informal version of Theorem~\ref{thm:scalar})
\label{res:informal}
The global minimum $\th^*$ of the non-convex objective~\eqref{eq:objintro} can be computed in three steps: \\
(i) Given the observations $\obs(k)$, compute the pseudo-observations $z(k)$ using~\eqref{eq:sym3}. \\ (ii) Using these pseudo-observations, estimate  the pseudo parameters  $\lam_1 = \th_1 + \th_2 + \th_3$, $\lam_2 = \th_1 \th_2 + \th_1 \th_3 + \th_2 \th_3$, $\lam_3 =  \th_1 \th_2\th_3$. Clearly~\eqref{eq:objintro} is a stochastic convex optimization problem in  pseudo-parameters $\lam_1,\lam_2,\lam_3$. Let $\lam_1^*,\lam_2^*, \lam_3^*$ denote the estimates.
\\
(iii) Finally, solve the  polynomial equation $s^3 + \lambda_1^* s^2 + \lambda_2^* s + \lambda_3^*=0$. Then  the roots\footnote{Strictly speaking   $\th_1,\th_2,\th_3$ are  factors.  The  root is  the negative of a factor.}  are $\th^*$. Computing the roots of a polynomial is equivalent to computing the eigenvalues of the corresponding companion matrix (Matlab command {\tt roots}).
\end{result}

Put simply the above result says that while~\eqref{eq:objintro} is non-convex in the roots of a polynomial, it is convex in the coefficients of the polynomial!
To explain Step (ii),
 clearly~\eqref{eq:objintro} is convex in the pseudo-parameters $\lam_1,\lam_2,\lam_3$.
We can straightforwardly compute the global minimum in terms of these pseudo parameters as $\lam_1^* = \E\{z_1\}$, $\lam_2^* = \E\{z_2\}$, $\lam_3^* = \E\{z_3\}$.

To explain Step (iii) of the above result, we use a crucial property of symmetric functions. The reader van verify that the following monic polynomial in variable $s$ satisfies
$$ (s+ \th_1) (s+\th_2) (s+\th_3)
= s^3 + \lambda_1 s^2 + \lambda_2 s + \lambda_3
$$
The above equation states  that a monic polynomial with pseudo-parameters $\lam_1,\lam_2,\lam_3$ as coefficients has the parameters $\th_1,\th_2,\th_3$ as roots of the polynomial. By the fundamental theorem of algebra, there is a unique invertible map between the coefficients of a monic polynomial and the set of roots of the polynomial.
As a result having computed the global minimum $\lam^*$ of the above objective~\eqref{eq:objintro}  (since it is convex in $\lam$), we can compute the unique parameter set $\th^*$, which are the set of roots of the corresponding polynomial. Thus we have computed  the global minimum $\th^*$
of the non-convex objective~\eqref{eq:objintro}. To summarize Result~\ref{res:informal} gives   a constructive method  to estimate the true parameter set $\tht$ given anonymized observations (albeit in an extremely simplified setting).

\subsection{Main Results and Organization}
\begin{compactenum}
  \item Our first main result in Sec.~\ref{sec:scalar},   extends the above simplistic formulation to a random  input process  $\reg(k)$ rather than a constant. To achieve this,  Theorem~\ref{thm:scalar} exploits the homogeneous property of the symmetric transform $\sym$ to construct a consistent estimator for $\tht$. 
In Theorem~\ref{thm:scalar}, we will construct a stochastic gradient algorithm that generates a sequence of estimates $\lam(k)$ that provably converges to $\lam^*$ (since the problem is convex). The roots of the corresponding polynomial converge to $\th^*$.

  \item Sec.~\ref{sec:vector}  extends this symmetric transform  approach to the case where  each anonymized observation $\obs_l(k)$ is a vector in $\reals^\thdim$ where 
 $\thdim\geq 2$ in~\eqref{eq:obs}.
For this vector case, three issues  need to be resolved:
\begin{enumerate}
\item It is not possible to use the  scalar symmetric transform~\eqref{eq:sym3} element-wise   on vector observations. Naively applying the scalar symmetric transforms  element wise   yields ``ghost'' parameters estimates that are jumbled across the various stochastic systems (see  Sec.\ref{sec:naive}.)
  \item Since a scalar symmetric transform  (or equivalently, the  one variable polynomial transform) is not useful, we will use a two-variable polynomial transform inspired by~\citet{MN93}. However, a new issue arises.
In the scalar observation case, we use the fundamental theorem of algebra to construct a unique mapping between the roots of a  polynomial and the coefficients of the  polynomial.
Unfortunately,  in general the fundamental theorem of algebra  does  not extend to polynomials in two variables. The key point we will exploit below is that the ring of two-variable polynomials is a unique factorization domain over the ring of one-variable polynomials.   This gives us a constructive method  to extend Theorem~\ref{thm:scalar} to sets of vector observations ($\thdim\geq 2$). This is the content of our main result Theorem~\ref{thm:vector}.
\item The final issue is that of homogeneity of the symmetric transform.
In the scalar case, the homogeneity property is crucial in the proof of Theorem~\ref{thm:scalar}. We construct a suitable multidimensional generalization for the vector case in order to prove Theorem~\ref{thm:vector}.
\end{enumerate}

\item {\em Asymptotic Covariance of Adaptive Filtering  Algorithm}:  
Sec.~\ref{sec:convergence}   analyzes the convergence and asymptotic covariance of the adaptive filtering algorithm~\eqref{eq:lmsvector}. In the stochastic approximation literature \citep{BMP90,KY03}, the asymptotic rate of convergence is specified in terms of the 
asymptotic covariance of the estimates. 
We study the asymptotic efficiency of the proposed  adaptive filtering  algorithm. Specifically we address the question: \textit{How much larger is the asymptotic covariance due to use of the symmetric transform to circumvent anonymization, compared to the classical LMS algorithm when there is no anonymization}?

\item {\em  Mixture  Model for Noisy Matrix Permutations}: We can assign a probability law  to the permutation process $\perm$ in the
  anonymized observation model~\eqref{eq:obs}, \eqref{eq:setobs} as follows:
   \begin{equation}
   \label{eq:hmmobs}
   \underset{\numtarget \times \thdim}{ \obs(k)} =  \underset{\numtarget\times\numtarget}{\perm(\state(k))}\,
   \underset{\numtarget\times \thdim}{\vphantom{\obs(k)}\tht}\, \underset{\thdim\times\thdim}{\reg(k)} + \underset{\numtarget\times \thdim}{\perm(\state(k))\,\err(k)}
 \end{equation}
 Here $\perm(\state(k))$ denotes a randomly chosen $\numtarget\times \numtarget$ permutation matrix that evolves according to some random process~$\state$.
 So~\eqref{eq:hmmobs} is a probabilistic mixture model.
The matrix valued observations $\obs(k)$  are  random permutations of the rows of matrix $\tht \reg(k)$ corrupted by noise. Given these observations, the aim is to estimate the matrix~$\tht$. Note that there are $\numtarget!$ possible permutation matrices $\perm$.


In the context of mixture models, Section~\ref{sec:mle} and Appendix~\ref{sec:em}  present  two results:\\
(i) {\em Mean-preserving Blackwell dominance and Anonymity of permutation process:}
Section~\ref{sec:mle}  uses the error probability  of  the Bayesian  posterior estimate of the random permutation state $\state(k)$ in~\eqref{eq:hmmobs}  as a measure of  anonymity. This is in line with~\citet{THG19} where anonymity is  studied in the context of mutual information and error probabilities. We will then use  Blackwell dominance and a novel result in mean preserving spreads  to relate this anonymity to the covariance of our proposed  adaptive filtering  algorithms. 

(ii) {\em Recursive Maximum likelihood estimation of $\tht$}
In Appendix~\ref{sec:em}, we discuss a recursive maximum likelihood estimation (MLE)  algorithm for
 the parameters $\tht$.  This requires knowing the density of $\err$ and the mixture probabilities  (of course these can be estimated, but given the $\numtarget!$ state space dimension, this becomes intractable).   A more serious issue  is that  the likelihood is not necessarily concave in $\th$. In comparison, our  symmetric function approach yields  a  convex stochastic optimization problem. 

\end{compactenum}

\subsection{Applications of Anonymized Observation Model}\label{sec:applications}

 We classify  applications of the anonymized observation model~\eqref{eq:obs}, \eqref{eq:setobs} into two types: (i) Due to sensing limitations,  the sensor  provides noisy measurements from multiple processes, and there is uncertainty as to  which measurement came from which process and (ii) examples where the identities of the processes generating the measurements are purposefully hidden to preserve anonymity.

 {\em 1. Sensing/Tracking Multiple Processes with Unlabeled Observations}:  The classical observation model  comprises a sensor (e.g.\ radar) that generates noisy measurements where, due to sensing limitations, there is uncertainty in the origin of the measurements. The observations are unlabeled and not assigned to a specific target process~\citep{BLK08}.
 In this context, estimating the underlying parameter $\tht$  of the target processes is identical to our estimation objective. As mentioned earlier,
 data association is widely studied in  Bayesian estimation for target tracking.  In this paper we focus on stochastic optimization with anonymized observations. for example, to estimate the underlying parameters, or more generally adaptively optimize a stochastic system comprising $\numtarget$ parallel process.

{\em 2. Adaptive Estimation with $k$-Anonymity and $l$-diversity}: We now discuss examples where the  labels (identities) of the $\numtarget$ processes are purposefully hidden.
Anonymization of trajectories arises in several applications including  health care where wearable monitors  generate time series of data uniquely matched to an individual, and connected vehicles, where location traces are recorded over time.

 The concept of $k$-anonymity\footnote{Data anonymity  is  mainly studied under two categories: $k$-anonymity and differential privacy. Differential privacy methods typically add noise to trajectory data providing a provable privacy guarantee for the data set. Even though we consider Laplacian noise for $\err$ in the numerical studies and this can be motivated in terms of differential privacy; we will not  discuss differential privacy in this paper.}  (we will call this  $\numtarget$-anonymity since we use $k$ for time) was proposed by~\citet{Swe02}.  It  guarantees that there are at least $\numtarget$ identical records in a data set that are indistinguishable.
In our formulation, due to the  anonymization step~\eqref{eq:setobs}, the identities (indexes) $\target$ of the $\numtarget$ processes are indistinguishable. 
More generally, 
in the model~\eqref{eq:obs}, \eqref{eq:setobs}, 
the  identity $l$ of each target itself can be a categorical vector $[l_1,\ldots,l_N]$.
For example if each process models GPS data trajectories of individuals, the categorical data $\reg_\target(k)$ records discrete-valued variables such as individuals identity, specific locations visited, etc. 
To ensure  $\numtarget$-anonymity, these categorical vectors are all allocated a  single vector, thereby  maintaining anonymity of  the categorical data. Thus the analyst only sees the anonymized observation set $\obs(k)$.

Note that $L$-anonymity hides identity $\target$ but discloses attribute information, namely the noisy observation set  $\obs(k)$.   To enhance $L$-anonymity, the attributes in $L$-anonymized data are often $M$-diversified\footnote{The  terminology used in the literature is ``$l$-diversified''; but we use $l$ for the index of the target process.} ~\citep{MKG07}:  each equivalence class  is constructed so that there are at least $M$ distinct parameters. In our notation, if at least $M$ processes have distinct  parameter vectors $\th_\target$, $\target = 1,\ldots,M$, then $M$-diversity of the attribute data is achieved.  

In our formulation,  the input signal matrices $\reg(k)$ are  the same for all $\numtarget$ processes. Thus the input matrices   also preserve $\numtarget$-anonymity.
If the analyst could specify a different input signal $\reg_\target$  to each system $\target$, then the analyst can straightforwardly estimate  $\tht_\target$ for each target process $\target$,
thereby breaking anonymity; see Remark 6 after Theorem~\ref{thm:scalar} below.

 {\em 3. Product Sentiment given Anonymized Ratings}: Reputation agencies such as Yelp post anonymized ratings or products. Market analysts aim to estimate the true sentiment of the group of users given these anonymized ratings \cite{LTE19}.

{\em 4. Evaluating Effectiveness of Teaching Strategy given Anonymized Responses}:
A teacher instructs $\numtarget$  students with input signal $\reg(k)$. Each student $\target$ has prior knowledge $\tht_\target$. and responds to the teaching input with answer
 $\obs_l(k)$.  The identity $\target$ of the student is hidden from the teacher.  Based on these anonymous responses, the teacher aims to estimate the students prior knowledge $\tht$.  See also~\cite{LMR21} for other examples. Anonymized trials are also used in evaluating the effectiveness of drugs vs placebo.


 \section{Adaptive Filtering with scalar anonymized observations}
 \label{sec:scalar}
 For ease of exposition,
 we first discuss   the problem of estimating the true parameter  $\tht$ when the observation $\obs_\target(k)$ of each process $\target$ is a scalar; so $\thdim = 1$
 in~\eqref{eq:obs} and $\reg(k)$ is a scalar.
 Since there are 
 $\numtarget $ independent  scalar processes in~\eqref{eq:obs}, the parameters generating these $\numtarget$ processes  is $\tht = \{\tht_1,\ldots,\tht_\numtarget\}$.

 Given  the anonymized observation set  $\obs(k)=\{\obs_1(k),\ldots,\obs_\numtarget(k)\}$ at each time $k$, our main idea is to  construct a pseudo-measurement vector $\z(k) \in \reals^\numtarget$. Suppressing the time dependency $(k)$ for notational convenience, we  construct the $\numtarget $ pseudo-measurements
 $z_l, l \in \numtargetset$ via a symmetric transform\footnote{By symmetric transform $\sym_\target$,  we mean
   $\sym_\target\{\obs_1,\ldots,\obs_\numtarget\} = \sym_\target \{P \cdot \{\obs_1,\ldots,\obs_\numtarget \} \}$ for any permutation $P$ of
   $\{\obs_1,\ldots,\obs_\numtarget\}$. Thus while the elements  $\{\obs_1,\ldots,\obs_\numtarget\}$ are arbitrarily ordered, the value of 
   $\sym_\target\{\cdot\}$ is unique. Eq~\eqref{eq:polydef} gives a systematic construction of such symmetric transforms that is uniquely invertible, see~\eqref{eq:ipoly}.}  \citep{Mac98}  
 as follows:
\begin{equation} \label{eq:pseudo}
  \begin{split}
      z = \sym\{\obs\}  \iff    z_\target = \sym_\target\{\obs_1,\ldots,\obs_\numtarget\}  \\ \defn \sum_{i_1< i_2<\cdots < i_\target} \obs_{i_1}\,\obs_{i_2} \cdots \obs_{i_\target},\; l \in \numtargetset 
 \end{split} 
\end{equation}
Recall our notation $\numtargetset = \{1,\ldots,\numtarget\}$.
It is easily shown using the classical Vieta's formulas~\citep{Jac12}, that 
the pseudo-measurements  $z_l, l \in \numtargetset$ in~\eqref{eq:pseudo} are the coefficients of the following $\numtarget$-order polynomial in variable~$s$:
\begin{equation}
  \label{eq:polydef}
 \sym\{\obs\}(s) \defn  \prod_{\target=1}^\numtarget (s+\obs_\target) =s^\numtarget +     \sum_{\target=1}^\numtarget z_{\target}\,s^{\numtarget-\target}
\end{equation}

As an  example, consider  $\numtarget =3 $ independent scalar processes. Then  the pseudo-observations using~\eqref{eq:pseudo} are given by~\eqref{eq:sym3}.
The reader can verify that the pseudo-observations $z_1,z_2,z_3$ are  the coefficients of the polynomial
$(s+\obs_1)(s+\obs_2) (s+\obs_3)$.

Note that each $z_l$ is permutation invariant:  any permutation of the elements of $\{\obs_1,\ldots,\obs_\numtarget\}$  does not affect~$z_l$. That is why our notation above involves the set
$\{\obs_1,\obs_2,\ldots, \obs_\numtarget\}$.  

{\em Remark}: It is easily verified  from~\eqref{eq:pseudo} that  the symmetric transforms $\sym_\target$ is homogeneous of degree $\target$:  for any $c\in \reals$,
\begin{equation}
  \label{eq:scale}
\sym_\target\{c\,\th_1,\ldots,c\,\th_\numtarget\} = c^\target \, \sym_\target\{\th_1,\ldots,\th_\numtarget\},  \quad\target \in \numtargetset
\end{equation}

\subsection{Symmetric Transform and Estimation Objective}  
Given the set valued sequence of anonymized observations, $\obs(1),\obs(2), \ldots \obs(k),\ldots$ generated by~\eqref{eq:obs},  our aim  is to estimate  the true parameter set $\tht=\{\tht_1,\ldots,\tht_\numtarget\}$.
To do so,
we first construct the pseudo measurement vectors  $z(1),z(2),\ldots, z(k)$ via~\eqref{eq:pseudo}.
Denoting $\th = \{\th_1,\ldots,\th_\numtarget\}$,  our objective is to estimate  the set $\th^*=\{\th_1^*,\ldots,\th_\numtarget^*\}$
that  minimizes: 
\begin{equation}
  \label{eq:multiobj0}
  \begin{split}
  \th^*=     \argmin_\th \sum_{\target \in \numtargetset} \E\ | z_\target - \sym_\target\big\{\reg \,\th_1, \reg\, \th_2,\ldots, \reg \, \th_\numtarget\big\} |^2\\
    \text{where } z_\target = \sym_\target\big\{\reg\, \tht_1 + \err_1, \ldots,  \reg\, \tht_\numtarget + \err_\numtarget \big\}                \end{split}
\end{equation}
Recall the symmetric transform $\sym_\target$ is defined in~\eqref{eq:pseudo}.
Finally, define the symmetric transforms on the model parameters as
\begin{equation}
  \label{eq:lam}
\lam = \sym\{\th\} \iff  \lam_\target = \sym_\target\{\th_1,\ldots,\th_\numtarget\},\;\target \in \numtargetset.
\end{equation}
Note that  $\lam=[\lam_1,\ldots,\lam_\numtarget]^\p$ is an $\numtarget$-dimension vector whereas $\th$ is a set with $\numtarget $ (unordered) elements.

From~\eqref{eq:multiobj0}, we see that $\th^*$ is a second order method of moments estimate of $\tht$ wrt pseudo observations. Importantly, this  estimate is independent of  the anonymization map $\perm$.

\subsection{Main Result. Consistent Estimator for $\tht$.}
  We are now ready to state our main result,  namely an adaptive filtering  algorithm to estimate $\tht$ given anonymized scalar observations.     The  result says that while objective~\eqref{eq:multiobj0} is non-convex in  $\th$, we can reformulate it as a convex optimization problem in terms of $\lam$ defined in~\eqref{eq:lam}. The intuition is that  the objective~\eqref{eq:multiobj0} is non-convex in the roots of the polynomial (namely, $\th$) , but is convex in the {\em coefficients} of the polynomial (namely, $\lam$); and by the fundamental theorem of algebra there is a one-to-one map from the coefficients $\lam$ to the roots $\th$.  Therefore, by mapping observations to pseudo observations (coefficients of the symmetric polynomial),  we can construct a globally optimal estimate of~\eqref{eq:multiobj0}. 
  
\begin{theorem} \label{thm:scalar}
  Consider the sequence of  anonymized observation sets $(\obs(k), k \geq 1)$ generated by~\eqref{eq:obs} and \eqref{eq:setobs}, where $\reg(k)$ is a known iid scalar sequence. Then
  \begin{enumerate}
    \item The objective~\eqref{eq:multiobj0} can be expressed as
  $\numtarget $ decoupled convex  optimization problems in terms of  $\lam$ defined in~\eqref{eq:lam}:
  \begin{equation} \label{eq:decoupled}
    \begin{split}
    \min_{\lam_\target}  \E| z_\target - \reg^l  \lam_l|^2 \quad
    \text{ where } \\   \quad
      z_l  (k)     = \big(\reg(k)\big)^l\, \lamt_\target + w_\target (k)
    \end{split}
  \end{equation}
   The process $w(k)$ is defined explicitly in~\eqref{eq:expansion} below.
  \item 
    The global minimizer $\th^*$ of objective~\eqref{eq:multiobj0} is consistent  in the sense that $\th^* = \tht$. 
  \item\label{item:1}
With pseudo observations $z(k) = \sym\{\obs(k)\}$ defined in~\eqref{eq:pseudo},
consider the  following bank of $\numtarget$ decoupled adaptive filtering  algorithms operating on  $z(k)$:  Choose $\lam(0) \in \reals^\numtarget$. Then
for $\target \in \numtargetset$, update as
      \begin{equation}
        \label{eq:lmsbank}
        \begin{split}
    \lam_\target(k+1) &= \lam_\target(k) + \ep \, \reg^\target(k)\, \big(z_\target(k)  - \reg^\target(k)\, \lam_\target(k)\big)    \\
          \th(k+1) &= \Re\big(\sym^{-1}(\lam(k+1))\big)
        \end{split}
      \end{equation}
  Here $\sym^{-1}$ is defined in ~\eqref{eq:ipoly} and $\Re$ denotes the real part of the complex vector.
The estimates $\th(k)$ converge  in probability and mean square to $\th^*$ (see Theorem~\ref{res:kushner}).
\end{enumerate}
\end{theorem}

\subsubsection*{Discussion}  1.  Theorem~\ref{thm:scalar} gives a tractable and consistent  method for estimating the parameter set $\tht$ of the $\numtarget$ stochastic systems given set valued anonymized  observations $\obs(1),\obs(2),\ldots$. We emphasize  that  since the observations $\obs(k)$ are set-valued, the ordering of the elements of $\tht$ cannot be recovered; Statement  1 of the theorem asserts that  the set-valued estimate $\th^*$ converges to $\tht$.
Statement 2 of the theorem gives an adaptive filtering  algorithm~\eqref{eq:lmsbank} that operates on the pseudo observation vector  $z(k)$. Applying the transform $\sym^{-1}$ to the estimates $\lam(k)$  generated by~\eqref{eq:lmsbank} yields estimates $\th(k)$ that   converge   to the global minimum $\th^*$.
Since by assumption $\tht \in \reals^\numtarget$,  the second step of~\eqref{eq:lmsbank}  chooses the real part of the possibly complex valued roots.

2. An important property of the  symmetric operator $\sym$ is that it is  uniquely invertible  since   any $\numtarget$-th degree  polynomial has  a unique set of at most $\numtarget $ roots. Indeed,  given $\lam = \sym\{\th\}$,
$ \th = \sym^{-1}(\lam) $
are the unique set of roots $\{\th_1,\ldots,\th_\numtarget\}$ of the polynomial with coefficients $\lam_\target, \target \in \numtargetset$, that is,
\begin{equation}
    \label{eq:ipoly}
  \th = \sym^{-1}(\lam) \iff   s^\numtarget +     \sum_{\target=1}^\numtarget \lam_{\target}\,s^{\target-1} =
    \prod_{\target=1}^\numtarget (s+\th_\target)
  \end{equation}
  Note that $\sym^{-1}(\cdot)$ maps the vector $\lam$ to unique set $\th$. Recall
  that $\sym\{\cdot\}$ maps  set $\th$ to unique  vector $\lam$.
Computing the roots of a polynomial is equivalent to computing the eigenvalues of the  companion matrix e.g.,  Matlab command {\tt roots}. 

3. Typically the roots of a polynomial can be a sensitive function of the coefficients. However, this does not affect algorithm~\eqref{eq:lmsbank} since it operates on the coefficients only. The roots are {\em not} fed back iteratively  into algorithm~\eqref{eq:lmsbank}. In Section~\ref{sec:convergence} and Theorem~\ref{thm:sensitivity} below, we will quantify this sensitivity in terms of the asymptotic covariance of algorithm~\eqref{eq:lmsbank}.

4.  The adaptive filtering  algorithm~\eqref{eq:lmsbank} uses a constant step size; hence it converges weakly (in distribution) to the true parameter $\tht$~\citep{KY03}. Since we assumed $\tht$ is a constant, weak convergence is equivalent to convergence in probability. Later we will analyze the tracking capabilities of the algorithm when $\tht$ evolves in time according to a hyper-parameter.

5.  A stochastic gradient algorithm operating directly on objective~\eqref{eq:multiobj0} is
\begin{equation}
  \label{eq:sgdirect}
  \th(k+1) = \th(k) - \ep \, \nabla_\th \sum_{l \in \numtargetset} | z_\target(k)  
  - \sym_\target\big\{\reg(k) \,\th_1(k), ,\ldots, \reg(k) \, \th_\numtarget(k)\big\} |^2
\end{equation}
We show  via  numerical examples in Sec.\ref{sec:numerical} that objective~\eqref{eq:multiobj0} has  local minima and stochastic gradient algorithm~\eqref{eq:sgdirect}   can get stuck at these local minima. In comparison, the formulation involving pseudo-measurements yields a convex (quadratic) objective and algorithm
\eqref{eq:lmsbank} provably converges to the global minimum.   There is also another problem with~\eqref{eq:sgdirect}. If the initial condition  $\th(0)$ is chosen with equal elements, then since the gradient $\nabla_\th$ is symmetric (wrt $\obs$ and $\th$), all the elements of the estimate  $\th(k)$ have equal elements at each time $k$, regardless of the choice of $\tht$, and so algorithm~\eqref{eq:sgdirect} will not converge to~$\tht$.

{\em 6. Anonymization of input signal $\reg(k)$}:
We assumed that  the input signal matrices $\reg(k)$ are  the same for all $\numtarget$ processes. 
If the analyst can specify a different input signal $\reg_\target$  to each system $\target$, then the analyst can  estimate  $\tht_\target$ for each target process $\target$ via classical least squares, thereby breaking anonymity as follows:
Minimizing  $\E\{\sum_{\target \in \numtargetset} \obs_\target -  \reg_\target \th_\target\}^2 =
\E\{z_1 - \sum_{\target \in \numtargetset}   \reg_\target \th_\target\}^2
$ wrt $\th_\target $ yields the classical least squares  estimator.
Thus the analyst 
only needs the pseudo observations $z_1(k) = \sum_\target \obs_\target(k)$ to estimate $\tht_\target$
and thereby break anonymity.

In our formulation, since  the regression input signals $\reg_l$ are  identical,
minimizing $\E\{z_1 -  \reg\, \sum_{\target \in \numtargetset}   \th_\target\}^2$  only estimates   the sum of parameters, namely  $\sum_\target \tht_\target$; the individual parameters are not identifiable.  This is why  we require pseudo-observations $z_1,\ldots,z_\numtarget$  to estimate the  elements  $\tht_\target, \target \in \numtargetset$.

\section{Adaptive Filtering  given vector anonymized observations}  \label{sec:vector}
We now consider the case $\thdim \geq 2$, namely, for each process $\target \in \numtargetset$, the observation  $\obsl(k)$ in~\eqref{eq:obs}  is a $\thdim$-dimensional vector.
 We observe  the (unordered) set $\obs(k) = \{\obs_1(k), \ldots,\obs_\numtarget(k)\}$ at each time $k$. That is, we do not know which observation vector $\obs_\target(k)$ came from which process $\target$.
Given the anonymized observation set~\eqref{eq:setobs}, the aim is to estimate
$\tht \in \reals^{\numtarget\times \thdim}$.

{\em Remark}.
For each observation vector $\obs_\target \in \reals^\thdim$,
let $\obs_{\target,i}$ denote the $i$-th component. Note that  the elements of each vector $\obs_\target$ are ordered, namely $ \obs_\target= [\obs_{\target 1},\ldots,\obs_{\target \thdim}]^\p$, but the first index  $\target$ (identity of process) is anonymized yielding the observation set $\obs= \{\obs_1,\ldots,\obs_\numtarget\}$.

As mentioned in Sec.\ref{sec:intro}, 
for this vector case, three issues need to be resolved: First,
naively applying the scalar symmetric transforms  element wise   yields ``ghost'' parameter estimates that are jumbled across the various stochastic systems. (We discuss this in more detail below.)
 Second, we need a systematic way to encode the observation vectors via a symmetric transform that is invertible. We will use a two-variable polynomial transform. However, a new issue arises; in general the fundamental theorem of algebra, namely that an $\numtarget$-th degree polynomial has up to $\numtarget$ complex valued roots, does  not extend to polynomials in two variables. We  will construct an invertible map for two-variable polynomials.   This gives us a constructive method  to extend Theorem~\ref{thm:scalar} to vector observations $\thdim\geq 2$. 
The final issue is that of homogeneity of the symmetric transform.
Recall in the scalar case, the homogeneity property~\eqref{eq:scale}  was crucial in the proof of Theorem~\ref{thm:scalar}. We l need to  generalize  this to the vector case. The main result  (Theorem~\ref{thm:vector} below)
addresses these three issues.

\subsection{Symmetric Transform for Vector Observations}

This section constructs the 
symmetric transform $\sym$   for vector observations. The construction involves a polynomial in two variables, $s$ and $t$. It is convenient to first define the symmetric transform for an arbitrary  set  
$\al = \{\al_1,\ldots,\al_\numtarget\}$ where $\al_\target \in \reals^\thdim$.
The symmetric transform is defined as 
\begin{equation}
  \label{eq:multi}
  \begin{split}
    \sym\{\al\}(s,t) &=  
                        \prod_{\target=1}^\numtarget (s+ \sum_{i=1}^\thdim \al_{\target,i} \, t^{i-1} ) = s^\numtarget + \sum_{\target = 1}^\numtarget
\sum_{m=1}^{\M} \sym_{\target,m}\{\al\}\, s^{\target-1}\, t^{m-1} \\
\text{ where } & \M  \defn (\numtarget-\target)(\thdim-1) + \thdim    \end{split}
\end{equation}
So the symmetric transform is the array of polynomial coefficients $\sym_{\target.m}\{\al\}$
of the    above two variable polynomial. We write this notationally as 
$$ \sym\{\al\}  = [ \sym_{\target.m}\{\al\}, \;  m = 1,\ldots, \M , \; \target \in \numtargetset ]$$
When $\thdim=1$, we see that the symmetric transform~\eqref{eq:multi} specializes to~\eqref{eq:pseudo}. 

Another equivalent way of expressing the above  symmetric transform involves convolutions: The $\M$ dimensional vector  $S_l\{\al\} = [\sym_{\target 1}\{\al\},\ldots, \sym_{\target  \M}\{\al\}]^\p$ satisfies 
\begin{equation}
  \label{eq:convolution}
 \sym_\target\{\al\} = \sum_{i_1< i_2<\cdots < i_\target} \al_{i_1}\conv \al_{i_2} \conv \cdots \conv \al_{i_\target},  \quad \target \in \numtargetset
\end{equation}
where $\otimes$ denotes convolution.
Eq.~\eqref{eq:convolution} serves as a constructive computational method to
compute the symmetric transform of a set $\al$.

With the above  definition of the symmetric transform, consider  the observation set $\obs(k)= \{\obs_1(k), \ldots,\obs_\numtarget(k)\}$  at each time $k$. We define the pseudo-observations as 
\begin{equation}
  \label{eq:psuedomult}
  z(k) = \sym\{y(k)\}
\end{equation}

{\bf Example}. To illustrate the polynomial $\sym\{\obs\}(s,t)$,
 consider $\numtarget=2$ independent processes each of dimension $\thdim =2$. Then with $\obs_1 = [\obs_{11}, \obs_{12}]^\p$, $\obs_2 = [\obs_{21}, \obs_{22}]^\p$, the symmetric polynomial~\eqref{eq:multi} in variables $s,t$ is
\begin{equation}
  \label{eq:exmulti}
  \sym\{\obs\}(s,t) = (s+ \obs_{11} + \obs_{12} t) \, (s + \obs_{21} + \obs_{22} t) 
\end{equation}
Then the pseudo observations $z_{lm}$  specified  by the RHS of~\eqref{eq:multi} are the coefficients of this polynomial, namely
\begin{multline}
  \label{eq:psuedomulti}
  z_{11} = \obs_{11}\, \obs_{21}, \; z_{12} = \obs_{11}\, \obs_{22} + \obs_{12} \,\obs_{21} , \; z_{13} = \obs_{12} \,\obs_{22} ,\\  z_{21} = \obs_{11} + \obs_{21},  z_{22} = \obs_{12} + \obs_{22} 
\end{multline}
In the convolution notation~\eqref{eq:convolution}, the pseudo-observations are
$$z_1 =[z_{11},z_{12},z_{13}]^\p= \obs_1 \conv \obs_2 , \quad z_2 = [z_{21},z_{22}]^\p=\obs_1 + \obs_2 $$

We see from this example that the 
 pseudo-observations~\eqref{eq:psuedomulti} generated by the vector symmetric transform~\eqref{eq:multi}   is a superset of  the scalar symmetric transforms applied to each component of the vector observation.  Specifically
 pseudo-observations for the first elements  of $\obs_1$ and $\obs_2$, namely $\obs_{11},\obs_{2,1}$ are
 $z_{11}$, $z_{21}$. Similarly pseudo-observations or the second  elements  of $\obs_1$ and $\obs_2$, namely $\obs_{12}, \obs_{22}$  are
 $z_{13}$, $z_{22}$. But  $z_{12}$ in~\eqref{eq:psuedomulti}  is the extra pseudo-observation that cannot be obtained by simply constructing symmetric transforms of each individual element. In Sec.\ref{sec:naive} below, we will discuss the importance of the above vector symmetric transform compared to a naive application of scalar symmetric transform
 element-wise.

\subsubsection*{Why a naive element-wise symmetric transform is not useful} \label{sec:naive}
Instead of the vector symmetric transform defined in~\eqref{eq:multi}, why not  perform the scalar symmetric transform on each of the $\thdim$ components separately?  To make this more precise, let us define the \textit{naive} vector symmetric transform which uses the scalar symmetric transform $\sym_l$, $l \in \numtargetset$ in~\eqref{eq:polydef} as follows:
\begin{equation}
  \label{eq:naivesym}
 \bz_{\target j} =  \nsym_{\target,j}\{\obs\} =  \sym_{\target}\{\obs_{1,j},\ldots,\obs_{\numtarget,j}\} ,
  \quad j \in \{1,\ldots, \thdim\}
\end{equation}
This is simply the
scalar  symmetric transform $\sym\{\obs_{1 j}, \ldots,\obs_{\numtarget j}\}$ applied separately to  each component $j=1,\ldots,\thdim$.

In  analogy to~\eqref{eq:multiobj0}, we can define the estimation objective in terms of the naive vector transform as 
\begin{equation}
  \label{eq:naiveth}
  \bth^* = \argmin_\th \sum_{l \in \numtargetset} \sum_{j =1}^\thdim
  \E | \bz_{\target j} - \bsym_{\target,j}\{\reg \th_1, \reg \th_2, \ldots, \reg \th_\numtarget \}|^2
\end{equation}

The naive symmetric transform $\bsym$ in~\eqref{eq:naivesym}, \eqref{eq:naiveth} loses ordering information of the vector elements;  for example given two processes ($\numtarget =2 )$ each of dimension $\thdim = 2$,  $\bsym$ does not distinguish between observation set  $\{ [\obs_{1,1}, \obs_{1,2}] , [\obs_{2,1},\obs_{2,2}] \}$ and the observation set 
 $\{ [\obs_{1,1}, \obs_{2,2}] , [\obs_{2,1},\obs_{1,2}] \}$.
It follows that  $\bth^*$ in~\eqref{eq:naiveth} is not a consistent estimator for $\tht$; see remark following proof of Statement 4 of Theorem~\ref{thm:vector}.
Specifically, if the true parameters are $\tht = \{[\tht_{11}, \tht_{12}], [\tht_{21}, \tht_{22}]\}$, then the estimates can converge to the parameters of the ``ghost processes''
$\{ [\tht_{11}, \tht_{22}], [\tht_{21}, \tht_{12}]\}$.
That is,  the parameter estimates get jumbled between the stochastic systems.
Such ``ghost'' target estimates are common in data association in target tracking, and we will demonstrate  a  similar phenomenon in numerical examples of Sec.\ref{sec:numerical}
when using the naive symmetric transform on anonymized vector observations.

In comparison,  the vector symmetric transform~\eqref{eq:multi} systematically encodes the observations with no information loss. For example in the $\thdim=2, \numtarget=2$ case,  the extra pseudo-observation
$z_{12}$ in~\eqref{eq:exmulti} allows to distinguish between these observation sets.
(See also Appendix~\ref{sec:appendix} for the example $\thdim=3,\numtarget=3$.)
To summarize, the vector symmetric transform is fundamentally different to the scalar symmetric transform. 
We will use the vector symmetric transform as a consistent estimator for $\tht$ below.

\subsection{Main Result. Consistent Estimator for $\tht$}
We  first formalize our estimation objective based on the anonymized observations. Then we present  the main result.

Denoting $\th = \{\th_1,\ldots,\th_\numtarget\}$,  our objective is to estimate  the set $\th^*=\{\th_1^*,\ldots,\th_\numtarget^*\}$
that  minimizes the following  expected cost (where $\|\cdot\|_F$ denotes the Frobenius norm):
Compute
\begin{equation}
  \label{eq:multiobj2}
  \begin{split}
      \th^*=     \argmin_\th
    \E\| \sym\{\obs_1,\ldots,\obs_\numtarget\}  - \sym\{\reg \,\th_1, \reg\, \th_2,\ldots, \reg \, \th_\numtarget\}\big\|_F^2
  \end{split}
\end{equation}
Recall that $\th_\target \in \reals^\thdim$ for each $l \in \numtargetset$.
For notational convenience we use
$\{\reg \th\}$ to denote the set $ \{\reg \,\th_1, \reg\, \th_2,\ldots, \reg \, \th_\numtarget\big\} $. Also $\obs = \{\obs_1,\ldots,\obs_\numtarget\}$ is the (anonymized) observation set.

{\em Remark}. 
As in the scalar case, we note that  $\th^*$ in~\eqref{eq:multiobj2} is a second order method of moments estimate of $\tht$ wrt pseudo observations,   independent of anonymization map $\perm$.

We are now ready to state our main result, namely an adaptive filtering algorithm to estimate $\tht$ given the anonymized observation vectors.   As in the scalar case, the main idea is that we  have a convex optimization problem in the symmetric transform variables (denoted as $\lam$ below), and the  variables $\th$ can be evaluated by inverting the symmetric transform.
\begin{theorem}
  Consider the sequence of anonymized observation sets, $(\obs(k), k\geq 1)$  generated by~\eqref{eq:obs}, \eqref{eq:setobs}, where $\reg(k), k\geq 1$ is a known iid  sequence of $\thdim\times \thdim$ matrices. Then
  \begin{compactenum}
  \item 
The symmetric transform polynomial $\sym\{\obs\}(s,t)$  in~\eqref{eq:multi} can be decomposed into  signal and noise polynomials as
\begin{equation} \label{eq:multinoise}
  \sym\{\obs\}(s,t) = \sym\{\reg \tht\} (s,t) + w(s,t)
\end{equation}
where $w(s,t)$ is a noise polynomial whose coefficients are zero mean. (We define   $w(s,t)$ in~\eqref{eq:expansionm} below.)
\item The symmetric transform $\sym$ has the following homogeneity property:  With $\lam_{\target,m} \defn
  \sym_{\target,m}\{\th\}$,  then
  \begin{equation}
    \label{eq:vectorhom}
    \sym_{\target,m}\{\reg \th\} = \sum_{n \in \M} \lam_{\target, n}\, \sym_{\target,n}\{\reg^{\target,m}\} , \quad \target \in \numtargetset 
  \end{equation}
Here  for  $\lam_{m,n} = \sum_{i_1 \leq  i_2 \leq \cdots \leq i_l} \th_{1,i_1}\, \th_{2,i_2} \cdots \th_{l,i_l}$, we construct  $\reg^{\l,m}$ as the following $D \times l$ matrix of elements from input matrix $\reg$:
\begin{equation}
  \label{eq:regmat}
  \reg^{\target , m} \defn
  \begin{bmatrix}
    \reg_{i_1,1} & \reg_{i_2,1} & \cdots & \reg_{i_l,1} \\
    \reg_{i_1,2}  & \reg_{i_2,2}  & \cdots & \reg_{i_l,2} \\
    \vdots &  \vdots & \cdots  & \vdots \\
    \reg_{i_1,\thdim} & \reg_{i_2,\thdim} & \cdots & \reg_{i_l,\thdim} 
  \end{bmatrix}
\end{equation}
  
\item With pseudo observations $z = \sym\{\obs\}$ defined in~\eqref{eq:pseudo} and $\reg^{\target,m}$ defined in~\eqref{eq:regmat}, the  objective~\eqref{eq:multiobj2} can be expressed as $\numtarget$ decoupled convex optimization problems:
  \begin{equation}
  \label{eq:vectorsoln}
  \begin{split}
[\lam_{\target 1}^*, \ldots, \lam_{\target,\M}^*]&=  \argmin_{\lam_{l1},\ldots,\lam_{l\M }} \sum_{m \in \M} \E | z_{lm} \\ & \hspace{-2cm} - 
\sum_{n \in \M} \lam_{\target, n}\, \sym_{\target,n}\{\reg^{\target,m}\}|^2 
 \quad  l \in \numtargetset \\
  \th^*  &= \sym^{-1}(\lam^*)
  \end{split}
\end{equation}
\item The global minimizer $\th^*$ of objective~\eqref{eq:multiobj2} is consistent  in the sense that $\th^* = \tht$.

\item With pseudo-observations $z(k) = \sym\{\obs(k)\}$ computed by~\eqref{eq:convolution},
consider the  following  $\numtarget$ decoupled adaptive filtering  algorithms operating on   quadratic objective~\eqref{eq:vectorsoln}:  Choose initial condition  $\lam_{\target,m}(0) \in \reals$ arbitrarily.  Update each element  of $\lam_{\target.m}$, $ m \in \M, \target \in \numtargetset$ as
      \begin{equation}
        \label{eq:lmsvector}
        \begin{split}
          \lam_{\target,m}(k+1) &= \lam_{\target,m}(k) + \ep \,
                                  \sym_{\target, m}(\reg^{\target m}(k))  \\ 
      &   \hspace{-2cm} \times \sum_{m \in \M}  \big(  z_{\target,m}(k) - \sum_{n\in \M} \lam_{\target,n}(k) \,  \sym_{\target, n}\{\reg^{\target m}(k)\}   \big), 
\\
          \th(k+1) &= \Re\big(\sym^{-1}(\lam(k+1))\big)
        \end{split}
      \end{equation}
      Here $\ep>0$ is the algorithm step size,  $\sym^{-1}$ is evaluated via~\eqref{eq:thl1}, \eqref{eq:lineq},  $\reg^{l,m}$ is constructed in~\eqref{eq:regmat}, and $\sym_{\target,n}\{\cdot\}$ is computed in~\eqref{eq:convolution}.
Then       
 the estimates $\th(k)$ converge in probability and mean square to $\th^*$ (see Theorem~\ref{res:kushner}).

\end{compactenum}
\label{thm:vector}
\end{theorem}

\subsection{Discussion of Theorem~\ref{thm:vector}}

Despite the  complex notation, the important takeaway  from~\eqref{eq:vectorhom} is that
$\sym_{\target,m}\{\reg \th\} $ is a linear function of $\lam_{\target, n} = \sym_{\target,m}\{\th\}$. Therefore the objective~\eqref{eq:multiobj2} becomes a convex (quadratic) optimization problem~\eqref{eq:vectorsoln}. Thus similar to  the scalar case in Theorem~\ref{thm:scalar}, we have converted a non-convex problem in the roots  of a two-dimensional  polynomial to a convex problem in the coefficients of the polynomial.     Since the map between the set of roots and vector of coefficients  roots is uniquely invertible, the  optimization objectives~\eqref{eq:multiobj2} and~\eqref{eq:vectorsoln}  are equivalent.

\subsubsection*{Homogeneity of Symmetric Transform}
 The fundamental theorem of symmetric functions states that any symmetric polynomial can be expressed as a polynomial in terms of elementary symmetric functions~\citep[Theorem 4.3.7]{Sag01}.  However, Theorem~\ref{thm:vector}  exploits the linear map $\reg \th$ to obtain the specific result~\eqref{eq:vectorhom},  namely
 $ \sym_{\target,m}\{\reg \th\} = \sum_{n \in \M} \sym_{\target,n}\{\th\}\, \sym_{\target,n}\{\reg^{\target,m}\} $. This  qualifies as a vector version of  the homogeneity property~\eqref{eq:scale} in the scalar case.
 The scale factor is $\sym_{\target,n}\{\reg^{\target,m}\} $.

As a simple example of evaluating the matrix $\reg^{l.m}$ in~\eqref{eq:regmat}, suppose $\numtarget=3, \thdim=3$.
Then since $\lam_{11} = \th_{11} \th_{21} \th_{31}$, it follows from~\eqref{eq:regmat} and~\eqref{eq:multi} that
\begin{equation}
  \label{eq:exregmat1}
  \reg^{11} = \begin{bmatrix}  \reg_{11} & \reg_{11} & \reg_{11} \\
                \reg_{12} & \reg_{12} & \reg_{12}  \\
                \reg_{13} & \reg_{13} & \reg_{13}
              \end{bmatrix} , \;
              \sym_{11}\{\reg^{11}\}               = \reg_{11}^3, 
                      \end{equation}
$               \sym_{12}\{\reg^{11}\} = 3 \reg_{11}^2\reg_{12} $.
Also,  since $\lam_{12} = \th_{11} \th_{21} \th_{32} + \th_{11}\th_{22} \th_{31} + \th_{12} \th_{21} \th_{31}$, it follows from~that
\begin{equation}
  \label{eq:exregmat2}
  \begin{split}
  \reg^{12} = \begin{bmatrix}  \reg_{11} & \reg_{11} & \reg_{21} \\
                \reg_{12} & \reg_{12} & \reg_{22}  \\
                \reg_{13} & \reg_{13} & \reg_{23}
              \end{bmatrix} , \quad
              \sym_{11}\{\reg^{12}\}               = \reg_{11}^2\,\reg_{21}, \\
    \sym_{12}\{\reg^{21}\} = \reg_{11}^2\reg_{12} + \reg_{11} \reg_{12} \reg_{21} + \reg_{12} \reg_{11} \reg_{21}
  \end{split}
\end{equation}

\subsubsection*{$\sym$ is uniquely invertible}
The fundamental theorem of algebra, namely that an $\numtarget$-th degree polynomial has up to $\numtarget$ complex valued roots, does  not, in general, extend to polynomials in two variables. However, 
the above special construction which encodes the observations as coefficients of powers of $t$, ensures that $\sym$ is a uniquely invertible transform  between the set of observations and matrix of polynomial coefficients. This is because  the ring $F(s,t)$ of two-variable polynomials is a unique factorization domain over the ring $F(s)$ of one-variable polynomials \cite[Theorem 2.25]{Jac12}.

\subsubsection*{Evaluating $\sym^{-1}$}
Given the observations $\obs$, 
the transform $\sym\{\obs\}$  computes the pseudo-observations   via convolution~\eqref{eq:convolution}.
We now discuss how to compute $\th = \sym^{-1}(\lam)$ given $\lam$. This computation is required in~\eqref{eq:vectorsoln} to compute $\th^*$
and also in the adaptive filtering  algorithm~\eqref{eq:lmsvector} below.

As in the scalar case~\eqref{eq:polydef}, given
$\lam_{1,1},\ldots,\lam_{\numtarget,1}$,
we first compute $\th_{1,1}, \ldots, \th_{\numtarget,1}$ by
solving for the roots of the polynomial:
\begin{equation}
  \label{eq:thl1}
  \prod_{\target=1}^\numtarget (s+\th_{\target,1}) = s^\numtarget +     \sum_{\target=1}^\numtarget \lam_{\target,1}\,s^{\numtarget-\target}
\end{equation}
Next, solve for the remaining elements of $\th_{\target,m}$ iteratively over $m=2,3,\ldots,\thdim$. For each $m\geq 2$, 
given $\lam_{1,m},\ldots,\lam_{\numtarget,m}$ and $\{\th_{1,n},\ldots, \th_{\numtarget,n}\}$, $n=1,\ldots, m-1$,
we solve the following linear system of equations\footnote{It follows from the definition that $\sym_{i,m}$ is linear in $\th_{1,m},\ldots, \th_{\numtarget,m}$ with
  linear coefficients specified by  $\{\th_{1,n},\ldots, \th_{\numtarget,n}\}$, $n=1,\ldots, m-1$}
  for $\th_{1,m},\ldots, \th_{\numtarget,m}$:
  \begin{equation}
    \label{eq:lineq}
    \begin{split}
    S_{1,m}\{\th_{1,m},\ldots, \th_{\numtarget,m}\} &= \lam_{1,m} \\
      S_{2,m}\{\th_{1,m},\ldots, \th_{\numtarget,m}\} &= \lam_{2,m}     \\  &\vdots \\
 S_{\numtarget,m}\{\th_{1,m},\ldots, \th_{\numtarget,m}\} &= \lam_{\numtarget,m}   
    \end{split}
      \end{equation}
By the property of elementary symmetric polynomials, the linear system \eqref{eq:lineq} has full rank.

To summarize, computing $\sym^{-1}$ for the vector case  requires  solving a single polynomial equation  (as in the scalar case) and then $\thdim-1$ additional linear algebraic equations.

\subsection{Convergence of Adaptive Filtering Algorithm and Asymptotic Efficiency} \label{sec:convergence}

This section analyzes the convergence and asymptotic covariance of the adaptive filtering  algorithm~\eqref{eq:lmsvector}.
   The convergence  is typically studied via two approaches:  mean square convergence and weak convergence (since $\tht$ is assumed to be a constant, weak convergence to $\tht$ is equivalent to convergence in probability). We refer to the comprehensive books \citet{Say08,KY03,BMP90} for details.
   Below we state the main convergence result (which follows directly from these references). More importantly, we then discuss the asymptotic efficiency of the  adaptive filtering  algorithm~\eqref{eq:lmsvector}. Specifically we address the question: \textit{How much larger is the asymptotic covariance with the symmetric transform and anonymized observations, compared to the classical LMS algorithm with  no anonymization}?
   
The  algorithm~\eqref{eq:lmsvector} can be represented abstractly as
\begin{equation}
  \label{eq:sa}
  \lam(k+1) = \lam(k) + \ep \, \Reg(k)\, \big(z(k) - \Reg(k) \lam(k)\big)
\end{equation}
where $\Reg(k)$ is the block diagonal matrix $\diag(\sym_{\target m},\target \in \numtargetset, m \in \M)$.

Let ${\cal F}_k$ be the $\sigma$-algebra generated
by $\{ \Reg(n),\err(n), n<k, \lam(n),  n \le k\}$, and denote the conditional
expectation with respect to ${\cal F}_k$ by $\E_k$.
We assume  the following
conditions:

\begin{itemize}
\item[(A)] \label{A2} The signal $\{\Reg(k),\err(k)\}$ is independent of $\{\lam(k)\}$.
Either $\{\Reg(k),\err(k)\}$ is a
 sequence of bounded signals such that
there is a symmetric and positive definite
matrix $\Q$
such that $\E \Reg(k) \Reg^\p(k) =\Q$
\begin{equation}
  \label{eq:ineq-mix-1}
 \big\vert \sum^{\infty} _{n=k} \E_k [\Reg(n) \, \Reg'(n) - \Q] \big\vert
 \le K, \quad \big\vert \sum^{\infty} _{n=k} \E_k \Reg(k) e(n) \big\vert
 \le K , 
\end{equation}
 or $\{\Reg(k),\err(k)\}$
 is a sequence of martingale difference signals
satisfying $\E|\Reg(k)|^{4+\Delta}<\infty$ and $\E|\Reg(k) \err(k)|^{2+\Delta} <\infty$
for some $\Delta>0$.

\end{itemize}

Assumption A  includes correlated mixing processes
\citep[p.345]{EK86}.
and where 
the remote past and distant future are asymptotically
    independent. The boundedness is a mild restriction,
    for example, one may consider truncated processes. 
Practical implementations of stochastic gradient algorithms  use  a projection:
when the estimates   are
    outside a bounded set $H$, they are  projected back to the constrained set
    $H$. \cite{KY03} has extensively discusses  such projection algorithms.
    For
    unbounded signals,  (A) allows for  martingale difference sequences.

    \begin{theorem}[\citet{KY03}]  \label{res:kushner}
      Consider the adaptive filtering  algorithm~\eqref{eq:sa}. Assume (A). Then
        \begin{compactenum}
  \item  (Mean Squared convergence). For sufficiently large $k$,  the estimates $\lam(k)$ from adaptive filtering   algorithm~\eqref{eq:lmsvector} have mean square error
    $\E\{\|\lam(k) - \lamt\|^2\} = O(\ep)$. 

  \item (Convergence in probability) $\lim_{\ep\downarrow 0} P( \sup_{t \leq T} |\lam^\ep(t) - \lamt| > \eta) = 0$   as $T\rightarrow \infty$  for all $\eta> 0$.
Here $\lam^\ep(t) = \lam(k), \; t \in [\ep k, (\ep+1) k)$  denotes the continuous-time interpolated process  constructed from $\lam(k)$.
    
  \item (Asymptotic Normality). As  $k\rightarrow \infty$, for small $\epsilon$, the estimates $\lam(k)$ from  algorithm~\eqref{eq:lmsvector} satisfy the central limit theorem (where $\dist$ denotes convergence in distribution)
    \begin{equation}
      \label{eq:cltlam}
        {\ep}^{-1/2}\, \big( \lam(k) - \lamt  \big)  \dist  \normal\big(0, \Sig \big)
    \end{equation}
Here the asymptotic covariance $\Sig$ satisfies  the algebraic Lyapunov equation
  \begin{equation}
    \label{eq:Lyapunov}
    \Q \, \Sig + \Sig \,\Q  = R
  \end{equation}

\item (Asymptotic Covariance of Estimates). Therefore,  the estimates  $\th(k) = \sym^{-1}(\lam(k))$ satisfy
  \begin{equation}
    \label{eq:thcov}
    \begin{split}
    {\ep}^{-1/2}\, \big( \th(k) - \tht  \big)
    \dist \normal(0,\bSig), \\ 
      \bSig = \big(\nabla \sym^{-1} (\lamt)\big)^\p \,\Sig\, \nabla \sym^{-1} (\lamt) .
    \end{split}
  \end{equation}
\end{compactenum}
\end{theorem}

{\em Remarks}. (i) Statements 1,2 and 3 of the above result are well known \cite{KY03}. The expression for $\bSig$ in~\eqref{eq:thcov} follows from the  ``delta-method'' for asymptotic normality~\citep{Vaa00}. The delta-method requires that $\sym^{-1}$ is continuously differentiable. This  holds  since the solutions of a polynomial equation are continuously differentiable in the coefficients of the polynomial.  

(ii) Recall  $\th(k) = \sym^{-1}(\lam(k))$  is a set (and not a vector).  So we interpret~\eqref{eq:thcov} after ordering the elements in some specific way. In the scalar case, we can impose that the elements are  ascending ordered, namely, $\th_1\leq \th_2 \cdots \leq \th_\numtarget$. For the vector case, $\th$ can be ordered such that  the first elements of the parameter vector of the $\numtarget$ processes  are in ascending order, $\th_{11} \leq \th_{21} \leq \cdots \leq \th_{\numtarget 1}$.

(iii)  In the stochastic approximation literature \citep{BMP90,KY03}, the asymptotic rate of convergence is specified in terms of the 
asymptotic covariance of the estimates, namely $\Sig$ in~\eqref{eq:cltlam}  and $\bSig$
in~\eqref{eq:thcov}.
 Since we want to quantify the asymptotic convergence rate,
 we will focus on evaluating $\Sig$ and $\bSig$.

\subsection*{Loss in Efficiency due to Anonymity}

We  now  evaluate  the asymptotic  covariance matrices  $\Sig$ in~\eqref{eq:cltlam} and $\bSig$ in~\eqref{eq:thcov} to quantify the asymptotic rate of convergence of adaptive filtering algorithm~\eqref{eq:lmsvector}.
To obtain a tractable closed form expression, we consider  the scalar observation case $\thdim=1$.  So $\Sig$ and $\bSig$ are $\numtarget\times \numtarget$ covariance matrices. (Recall  there are $\numtarget$ anonymized processes.)

We assume that the zero mean noise process $\err(k)$ is iid across the $\numtarget$ processes with $\var\{\err_\target(k)\} = \sigma_\target^2$. Also we choose the regression input matrix as $\reg(k)
\sim \normal(0,I_{\numtarget\times \numtarget})$.
Using~\eqref{eq:lmsbank} it follows that for  $l \in \numtargetset$,
\begin{equation}
  \label{eq:Qval}
\begin{split}
  \Q &= \diag[\cov(\reg^{2l})] , \quad  \cov(\reg^{2l}) =  (2l-1) (2l-3) \cdots 1 
\end{split}
\end{equation}
Next define $\R_l = \cov[\reg^l(k) (\z_l(k) - \reg^l(k) \lam_l)]$ evaluated at $\lamt_l$.
We have
\begin{equation}
  \label{eq:Rval}
\R_l = \cov(\reg^l ( \reg^l (\lamt_l-\lam_l) + w_l)) \vert_{\lam=\lamt}= \cov(\reg^l w_l)
\end{equation}
These can be evaluated using the expression for $w$ in~\eqref{eq:expansion}.

Finally from Theorem~\ref{thm:sensitivity} in Appendix,  the sensitivity of the $l$-th polynomial root $\th_l$ wrt $m$-th coefficient $\lam_m$ is
\begin{equation}
  \label{eq:senspol}
  \begin{split}
  \nabla\sym^{-1}(\lam) &=  [\frac{d\th_l} {d\lam_m}],
\text{ where } 
 \frac{d\th_l} {d\lam_m}
=  \frac{(-1)^{m+1}\,(-\th_l)^{\numtarget-m} }{
    \frac{d\sym\{\th\}(-\th)}{d\th}\vert_{\th = \th_l}}
  \end{split}
\end{equation}
The above formula assumes that the polynomial does not have repeated roots; otherwise the sensitivity is infinite since $dS(-\th)/d\th = 0$ at a repeated root.

With the above characterization of $\Q, R, \nabla \sym^{-1} (\lam)$, we now evaluate   $\Sig$ and $\bSig$ explicitly for
$\numtarget=2$.

\begin{lemma}\label{lem:effloss}  Consider the anonymized model~\eqref{eq:obs}, \eqref{eq:setobs} with
 $\thdim=1$, $\numtarget =2 $. Assume the zero mean noise process $\err(k)$ is iid across the $\numtarget$ processes with $\var\{\err_\target(k)\} = \sigma^2$, and $\reg(k)
\sim \normal(0,I_{\numtarget\times \numtarget})$. Then the asymptotic covariance $\bSig$  (see~\eqref{eq:thcov}) of the estimates $\th(k)$ generated by algorithm~\eqref{eq:lmsbank} satisfies
\begin{equation}
\label{eq:tracesig}
\Tr(\bSig) = \frac{6 \sig^2 (\th_1^2 + \th_2^2) + \sig^4}{(\th_1 - \th_2)^2}
\end{equation}
\end{lemma}

{\em Remark}.
From~\eqref{eq:tracesig},    $\inf \Tr(\bSig) = 3 \sig^4$ when  $\th_1 = - \th_2 \rightarrow \infty$.
In comparison, for the classical LMS algorithm when the observations are not anonymized, the asymptotic covariance  for $\thdim=2$ is  $\Tr(\cov(\text{LMS}) )=  \sig^2$. So for $\thdim=2$, at best, the adaptive filtering algorithm~\eqref{eq:lmsbank} with anonymized observations is 3 times less efficient than the classical LMS.

\begin{proof}
From~\eqref{eq:Qval},
$ \Q = \begin{bmatrix} 1 & 0 \\ 0 & 3 \end{bmatrix} $.
Also~\eqref{eq:Rval} yields
$R =  \cov\begin{bmatrix} 
\reg (\err_1 + \err_2)  \\ \reg^2(\reg \err_2 \th_1 + \reg \err_1 \th_2 + \err_1 \err_2)
           \end{bmatrix} =\diag(2 \sig^2,  15\sig^2(\th_1^2 + \th_2^2) + 3 \sig^4)$.
Finally~\eqref{eq:senspol} yields    $ \nabla \sym^{-1}(\lam) = \begin{bmatrix} \frac{\th_1}{\th_1-\th_2} & \frac{\th_2}{\th_2-\th_1}
                                            \\  \frac{1}{\th_2-\th_1} & \frac{1}{\th_1-\th_2}
                                            \end{bmatrix} $.          
Then evaluating 
$\Sig = \frac{1}{2} \Q^{-1}\,R$, and $\bSig$ using~\eqref{eq:thcov} yields~\eqref{eq:tracesig}.
\end{proof}

To summarize, Lemma~\ref{lem:effloss} shows that for $\thdim=2$, at best, adaptive filtering with  anonymized data has three times  the asymptotic variance compared to the classical LMS algorithm.  

\subsection{Analysis for Tracking a Markov hyper-parameter}
So far we assumed that the true parameter $\tht$ was constant.     
An important property of a constant step size adaptive filtering algorithm~\eqref{eq:lmsvector} is the ability to track a time evolving true parameter. Suppose the true parameter $\tht(k)$ evolves according to a slow Markov chain with unknown transition matrix. How well does the adaptive filtering algorithm track (estimate) $\tht(k)$? Our aim is to quantify the  mean squared tracking error.

\begin{itemize}
\item[(B)] Suppose that exists  a small parameter $\mcstep>0$ and
 $\tht(k)$ is a discrete-time  Markov chain, whose
state space is
\begin{equation}
\label{eq:statespace}
 \M= \{ \hyper_1,\ldots,\hyper_m\}, \ \hyper_i
 \in\reals^{\numtarget \times \thdim}, i=1,\ldots,m,
\end{equation}
and whose transition probability matrix
$ P^\mcstep= I+ \mcstep Q$.
where $I$ is an
$\reals^{m\times m}$ identity matrix and $Q=(q_{ij})\in \reals^{m\times
m}$ is an irreducible
generator (i.e., $Q$ satisfies $q_{ij} \ge 0$ for $i\not =j$ and
$\sum^m_{j=1}q_{ij}=0$ for each $i=1,\ldots,m$)
of a continuous-time Markov chain.
\end{itemize}

The time evolving parameter $\tht(k)$ is called a hyperparameter. 
Although
 the dynamics of the hyperparameter $\tht(k)$  are used in our
analysis below,
the  implementation of the adaptive filtering
 algorithm~\eqref{eq:lmsbank}, does not use this information.

Define the tracking error of the adaptive filtering algorithm~\eqref{eq:lmsvector} as $\tilde{\lam}(k) \defn \lam(k)  -\lamt(k)$.
%
The aim is to determine 
 bounds on the tracking error $\tilde{\lam}(k)$ and therefore $\tilde{\th}(k)$.

\begin{theorem}
\label{thm:hyper}
Under  (A), (B), for sufficiently large $k$,
\begin{equation}
\label{eq:trackerror}
  \E |\tilde{\lam}(k) |^2 =  O(\ep + \mcstep + \mcstep^2/ \ep)
\end{equation}
Therefore, choosing $\mcstep = O(\ep)$, the mean squared-tracking error is $\E|\tilde{\lam}(k)|^2 = O(\ep)$ and so $\E|\tilde{\th}(k)|^2 = O(\ep)$ 
\end{theorem}
The proof follows from~\cite{YK05a}. The theorem implies  that even if  the  hyperparameter $\tht$  evolves on the same time scale (speed) as the adaptive filtering algorithm, the algorithm can   track the hyperparameter with mean squared error $O(\ep)$. 

 \section{Mixture Model for Anonymization} \label{sec:mle}
 This section uses a Bayesian interpretation of the anonymity map $\perm$ in~\eqref{eq:setobs} to present a performance analysis of the adaptive filtering algorithm~\eqref{eq:lmsvector}.
 Thus far we have assumed nothing about the permutation (anonymization) process $\perm$ in~\eqref{eq:setobs}. The  symmetric transform based algorithms proposed in  Sections~\ref{sec:scalar} and~\ref{sec:vector} are  oblivious to any assumptions on  $\perm$.
 Below we  formulate a probabilistic  model  for the permutation process  $\perm$.
 Based on this probabilistic model, we address two  questions:

 \begin{compactenum}
\item  \textit{How do noisy observations of the permutation process affect anonymity of the identity of the target processes?}
 We will consider the expected error probability  of the maximum posterior estimate of the permutation process as a measure of the anonymity of the permutation process.
 This  is in line with~\citet{THG19} where the error probability of an estimator  (and also mutual information) is used as a measure of anonymity.
 \item  \textit{How does anonymity of process $\perm$ in terms of Bayesian error probabilities relate to the asymptotic covariance of the adaptive filtering algorithm~\eqref{eq:lmsvector}?}
   Our main result below (Theorem~\ref{thm:blackwell}) shows  that if the observation likelihood of  noise process one Blackwell dominates that of noise process two, then  anonymity of the permutation process two  is higher than that of one;  and also the asymptotic covariance of the parameter estimates of the  adaptive filtering   algorithm~\eqref{eq:lmsvector} is higher.
 \end{compactenum}
 

 From a probabilistic point of view, the  anonymized observation model~\eqref{eq:obs}, \eqref{eq:setobs}  can be constructed as the following random permutation mixture model of the rows of matrix $\tht$:
 \begin{equation}
   \label{eq:probmodel}
   \underset{\numtarget \times \thdim}{ \obs(k)} = \perm(\state(k)) \, \obsv(k) =  \underset{\numtarget\times\numtarget}{\perm(\state(k))}\,
   \underset{\numtarget\times \thdim}{\vphantom{\obs(k)}\tht}\, \underset{\thdim\times\thdim}{\reg(k)} + \underset{\numtarget\times \thdim}{\perm(\state(k))\, \err(k)}
 \end{equation}
 Here $\perm(\state(k))$ denotes a randomly chosen $\numtarget\times \numtarget$ permutation matrix that evolves according to a random process
 $$\state\in \statespace, \quad  \statespace \subseteq \{1,2,\ldots,\statedim\} \text{ where } \statedim = \numtarget! $$
 since there are $\numtarget!$ possible permutations.
 Also $\obs(k) = [\obs_1(k),\ldots,\obs_\numtarget(k)]^\p$ where each 
 $\obs_\target(k) \in \reals^\thdim$.
 Recall that $\reg(k)$ is a known input (regression) matrix and  $\err(k) = [\err_1(k),\ldots,\err_\numtarget(k)]^\p$ is a $\numtarget \times \thdim$ matrix valued noise process whose elements  are zero mean.
 As previously, we assume for simplicity that $\err_\target(k)$ and $\err_m(k)$, $\target\neq m$  are iid vectors in $\reals^\thdim$.

\subsection{Anonymity of Permutation Process $\state$ and asymptotic covariance of adaptive filtering algorithm}
\label{sec:anonbayes}

This  section characterizes the anonymity of the permutation process $\state$ in terms of average error  probability of the \textit{maximum aposteriori} (MAP) state estimate.   
%
Our assumptions are:
\begin{compactenum}
\item 
The permutation process
$\state$ is iid with known  probabilities $\belief(i) \defn P(\state(k) = i) $.
\item The regression matrix $\reg(k) = I$. From a Bayesian  point of view, this is without loss of generality since $\reg(k)$ is known and invertible. So we can  post-multiply
\eqref{eq:probmodel} by $\reg^{-1}(k)$ to obtain an equivalent observation process.    
\end{compactenum}
Given the observation model~\eqref{eq:probmodel},
define the $\numtarget\times \thdim$-variate observation likelihood given   state $\state(k) = i$ as
\begin{equation}
  \label{eq:levels}
  \begin{split}
  \oprob_{i\obs} = \pdf(\obs(k) = \obs\, |\, \state(k) = i) \propto
    \pdf_\err\big(  \obs- \levels_i \big), \\
    \text{ where } \levels_i \defn \perm(i) \,\tht  \in \reals^{\numtarget\times \thdim}
  \end{split}
\end{equation}
Here $\pdf_\err$ denotes the $\numtarget\times \thdim$-variate density of noise process $\err$.
Since the $\numtarget$  noise processes  are independent, 
with $\obs_\target = \obs^\p \, e_\target$ where $e_\target \in \reals^\numtarget$ is the unit vector with 1 in the $\target$-th position, 
\begin{equation} \label{eq:bl}
  \oprob_{i\obs} =    \prod_{\target=1}^\numtarget \oprob_{i\obs_\target} ,
\oprob_{i\obs_\target} = \prod_{m=1}^\thdim \oprob_{i,\obs_{\target,m}}, \oprob_{i,\obs_{\target,m}}  = \pdf_{\err_{lm}}(\obs_{\target m} - \tht_{\target m})  
\end{equation}

The anonymity of the $\state$ depends on  the prior $\belief$ of the permutation process $\state$ and the observation likelihood $\oprob$.

{\em Perfect Anonymity}. 
If all $\statedim$ permutations are equi-probable, i.e., $\belief(i) = 1/\statedim$, then clearly
$P(\state(k)=i|\obs(k) ) = 1/\statedim$. So  the probability of error  of the maximum aposteriori estimate $\hat{\state}_k$ is  $P(\hat{\state}(k) \neq \state(k)) = (\statedim-1)/\statedim$ which is the largest possible value.
So for discrete uniform  prior on the permutation process,  perfect anonymity of the identities of the $\numtarget $ processes holds (even with no  measurement noise).

{\em Zero Anonymity}. If $\belief(\state) = 1$ for some  state $\state = i^*$, then the error probability is zero and there is no anonymity.

\subsubsection*{Anonymity of Permutation Process $\state$ wrt observation likelihood}
In the rest of this section, we analyze the anonymity of a Bayesian estimator of the permutation process $\state$ in terms of the observation likelihood $\oprob$, or equivalently, the noise $\err(k)$.
We start with Bayes formula for the posterior of permutation state $\state(k)$ given observation $\obs(k)$.
Define the diagonal matrix
$  \oprob_\obs = \diag[\oprob_{1\obs},\ldots,\oprob_{\statedim \obs}]$.
Then given the prior $\belief$ and observation $\obs(k)$,  the  posterior $\belief(k) = [\belief_1(k),\ldots\belief_\statedim(k)]^\p$ where $\belief_i(k)=\pdf (\state(k)=i\,\vert \,\obs(k))$ 
 is given by Bayes formula:
\begin{equation}
  \label{eq:bayes}
  \belief(k) = \filter(\belief,\obs(k)) \defn \frac{\oprob_{\obs(k)} \belief}{\filterd(\belief,\obs(k))}  ,  \text{ where }
    \filterd(\belief,\obs) = \ones^\p \oprob_{\obs(k)} \belief \end{equation}
Finally, given the posterior computed by~\eqref{eq:bayes},
define the maximum aposteriori (MAP)   permutation state estimate  as
$$\hat{\state}(k) = \argmax_i \belief_i(k) $$

\begin{lemma}
\label{lem:map}
The expected error probability  of the MAP state estimate is (where $\mathcal{Y}$ below denotes the observation space)
\begin{equation*}
\Pavg(\belief;\oprob)  = \E_\obs\{ P(\state(k) \neq \hat{\state}(k) | \obs) \} =1 - \inty  \max_i e_i^\p\oprob_{\obs}\, \belief  d\obs
\end{equation*}
where $e_l\in \reals^\statedim$ is the unit vector with 1 in the $i$-th position.
\end{lemma}

We normalize the expected error probability by defining the anonymity of permutation process $\state$  as
\begin{equation}
  \label{eq:anon}
\am(\belief,\oprob) = \Pavg(\belief;\oprob) \,\frac{\statedim}{\statedim-1}   \in [0,1]
\end{equation}
So  the anonymity $\am = 0$   when $\Pavg(\belief;\oprob) = 0$, and $\am = 1$
when $\Pavg(\belief;\oprob) = \frac{\statedim-1}{\statedim}$. 

\subsection{Blackwell Dominance and Main Result}
We now use  a novel result involving Blackwell dominance of mean preserving  spreads to relate the anonymity to the covariance of adaptive filtering algorithm~\eqref{eq:lmsvector}.

\begin{definition}[Blackwell ordering of stochastic kernels] We say that likelihood $\oprob$ Blackwell dominates likelihood $\boprob$, i.e.,  $\oprob \blackwell \boprob$ if $\boprob = \oprob \kerM$ where  $\kerM$  is a stochastic kernel. That is,
$\inty  \kerM_{\bar{y},y} \,dy = 1$ and $\kerM_{\bar{y},y} \geq 0$.
\end{definition}

Intuitively $\boprob$ is noisier than $\oprob$.
Thus 
observation $\obs$ with conditional distribution specified by $\oprob$ is 
said to be more  informative than (Blackwell dominates) observation $\bobs$ with conditional distribution $\boprob$,
see \cite{Kri16} for several applications.
When $\obs$ belongs to a finite set, 
it is well known~\citep{Sha58} that $\oprob_{\blackwell} \boprob$ implies that $\boprob$ has smaller Shannon capacity  than $\oprob$.

\subsubsection*{Main Result}

First we list the main assumptions:
\begin{enumerate}[label=(A\arabic*)]
\item \label{a:blackwell} $\oprob \blackwell \boprob$
\item \label{a:zeromean}  $\inty \oprob_{i\obs} \,\obs d\obs  = \levels_i$  and $\inty \boprob_{i\obs} \, \obs  d\obs= \levels_i$  (zero mean noise) 
\end{enumerate}
Recall $\levels_i$ is defined in~\eqref{eq:levels}.

Since the observations of the $\numtarget$ processes are independent, Blackwell dominance of the $\target$ individual likelihoods 
$\oprob_{iy_l} \blackwell \boprob_{iy_l}$, $\target\in \numtargetset$  is sufficient for~\ref{a:blackwell}.
The mean preserving spread assumption~\ref{a:zeromean} on $\oprob$ and $\boprob$ implies that the observation noise is zero mean. This  is a classical assumption for  the
convergence of the stochastic gradient algorithm~\eqref{eq:lmsvector}.


We are now ready to state the main result.
Theorem~\ref{thm:blackwell}  shows that  Blackwell ordering of observation likelihoods
yields an ordering for error probabilities (anonymity) and also  a  partial ordering on the asymptotic covariance matrices of the adaptive filtering  algorithm~\eqref{eq:lmsvector}.  So the more the anonymity of the permutation process, the higher the asymptotic covariance of the adaptive filtering  algorithm~\eqref{eq:lmsvector}. To the best of our knowledge, this result is new.

\begin{theorem} \label{thm:blackwell}
  Consider observations $\obs(k)$ generated by~\eqref{eq:probmodel}. 
  \begin{compactenum}
  
  \item $\cov_\oprob(\obs) \preceq \cov_{\boprob}(\obs)$ implies $\cov_\oprob(\sym\{\obs\})
    \preceq \cov_{\boprob}(\sym\{\obs\})$ for the symmetric transform $\sym$.

  \item Assume~\ref{a:blackwell}. Then the  average error probabilities satisfy $\Pavg(\belief;\oprob) \leq \Pavg(\belief;\boprob)$, and therefore the anonymity satisfies $\am(\belief,\oprob)
\leq \am(\belief,\boprob)$.
\item Assume ~\ref{a:blackwell}, \ref{a:zeromean}. Then $\cov_\oprob(\obs) \preceq \cov_{\boprob}(\obs)$.  Therefore,  the asymptotic covariance of $\lam(k)$ in~\eqref{eq:cltlam} of the adaptive filtering  algorithm satisfies 
  $\kalmancov(\oprob) \leq \kalmancov(\boprob) $. Also the asymptotic covariance
 of $\th(k)$ in~\eqref{eq:thcov}  satisfies $\bSig(\oprob) \leq \bSig(\boprob)$.
\end{compactenum}
\end{theorem}

The proof in the appendix uses  mean-preserving convex dominance  from Blackwell's classic paper~\citep{Bla53}.
Note that Theorem~\ref{thm:blackwell} does not require the  noise to be  Gaussian; for example, the noise can be  finite valued random variables.

To summarize, we have linked anonymity of the observations (error probability of the Bayesian MAP estimate) to the asymptotic covariance (convergence rate) of the adaptive filtering algorithm~\eqref{eq:lmsvector}.

\section{Numerical Examples}\label{sec:numerical}

\subsubsection*{Example 1: Symmetric Transform for Scalar case $\thdim=1$}
The aim of this example is to show that  objective~\eqref{eq:multiobj0}  has local minima wrt $\th$; and therefore the classical stochastic gradient algorithm~\eqref{eq:sgdirect} gets stuck in a  local minimum. In comparison, the objective~\eqref{eq:decoupled} in terms of pseudo-measurements is   convex  (quadratic) wrt  $\lam$ and therefore the adaptive filtering algorithm~\eqref{eq:lmsbank} converges to the global minimum $\th^*$.
   
We consider $\numtarget=3$ independent scalar processes ($\thdim=1$)
with anonymized observations generated  as in~\eqref{eq:setobs}.
The true model that generates the observations is $\tht=[-2,\; 5,\; 8]^\p$. 
The regression signal $\reg(k) \sim \normal(0,\sigma^2)$ where $\sigma=1$.
The noise error $v(k) \sim \normal(0,\sigma_v^2)$ where $\sigma_v = 10^{-2}$.

\begin{figure*}[p]
  \centering
  \begin{subfigure}{0.4\linewidth}
    \includegraphics[scale=0.4]{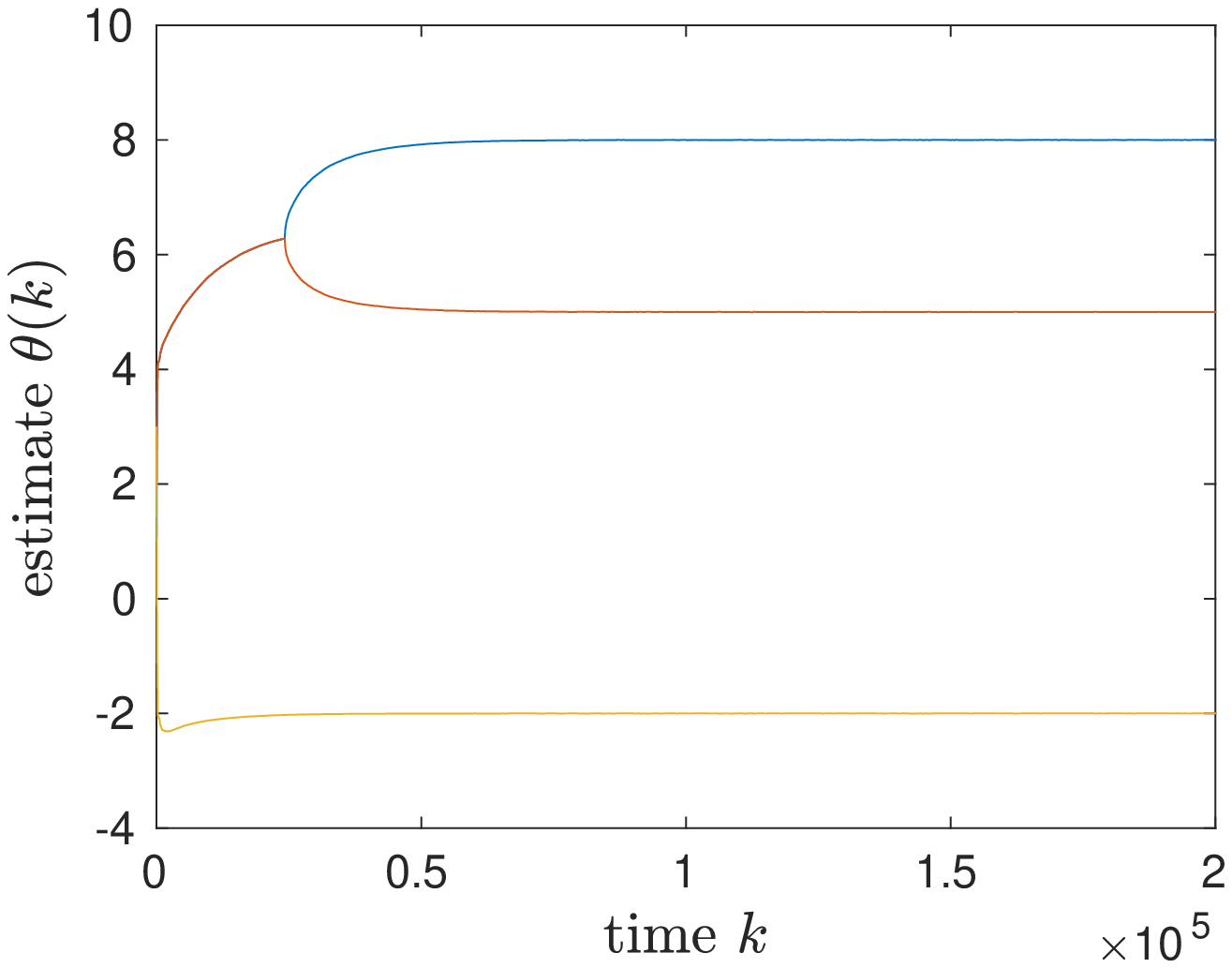}
    \caption{Algorithm~\eqref{eq:lmsbank} converges to global optimum $\tht$.}
    \label{fig:onedimlam}
  \end{subfigure} \hspace{1cm}
  \begin{subfigure}{0.4\linewidth}
    \includegraphics[scale=0.4]{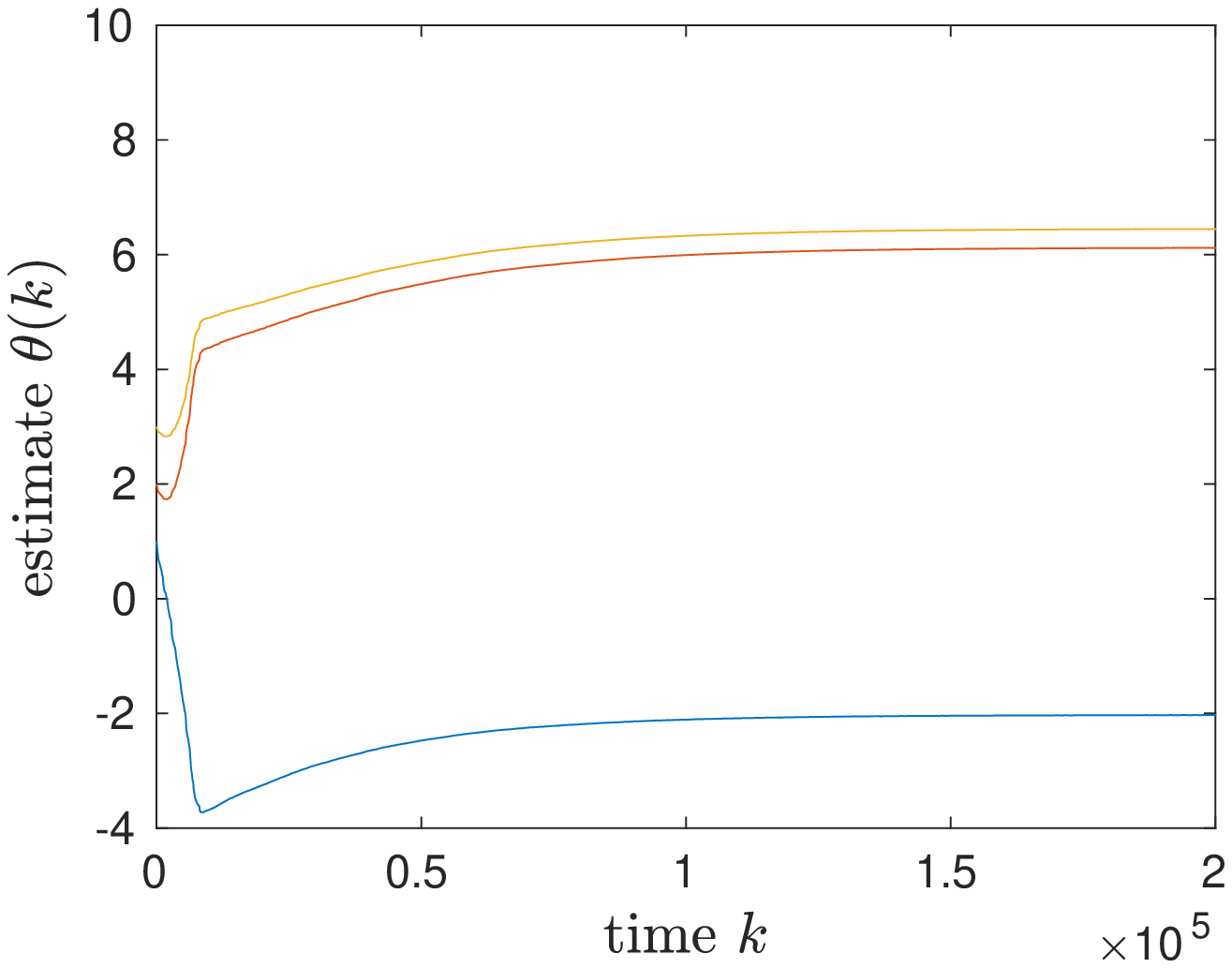}
    \caption{Classical stochastic gradient algorithm~\eqref{eq:sgdirect}  gets stuck in local minimum.}
    \label{fig:onedimth}
  \end{subfigure}
  \caption{Anonymized estimation problem in Example 1 of Sec~\ref{sec:numerical}.  The initial condition is $\th(0) = [1,2,3]^\p$ and
    the true parameter is $\tht = [-2, 5, 8]^\p$. 
    Fig.\ref{fig:onedimlam} shows that the parameter estimates generated by Algorithm~\eqref{eq:lmsbank} converge to  $\tht$.   Fig.\ref{fig:onedimth} shows that the parameter estimates generated    by  stochastic gradient algorithm~\eqref{eq:sgdirect} operating on~\eqref{eq:multiobj0}  do not converge to~$\tht$. }   
  \label{fig:single}
\end{figure*}

We ran the adaptive filtering algorithm~\eqref{eq:lmsbank} on a sample path of $2\times 10^5$  anonymized observations generated by the above model with
 step size  $\ep = 10^{-4}$. For initial condition  $\th(0) = [1,2,3]^\p$, Figure~\ref{fig:onedimlam}  shows  that the estimates generated by Algorithm~\eqref{eq:lmsbank} converges to $\tht$.
As can be seen from Figure~\ref{fig:onedimlam}, the sample path of the estimates
initially are coalesced,  and then split. This is because the estimates $\th_1(k)$ and $\th_2(k)$ are initially complex conjugates; since we plot the real parts, the estimates of $\th_1(k)$ and $\th_2(k)$ are identical.  

We also ran  the classical stochastic  gradient   algorithm~\eqref{eq:sgdirect} on the anonymized observations. Recall this algorithm minimizes~\eqref{eq:multiobj0} directly. The step size chosen was $\ep=10^{-7}$   (larger step sizes led to instability).  For initial condition  $\th(0) = [1,2,3]^\p$, Figure~\ref{fig:single}(b) shows that the estimates converge to a local stationary point $[-2.02, 6.12, 6.45]^\p$ which is  not $\tht$.
On the other hand for initial condition $\th(0) = [3,6,9]^\p$, we found that  the estimates converged
to $\tht$. This provides numerical verification that objective~\eqref{eq:multiobj0} is non-convex.
Besides the non-convex objective, another problem with the algorithm~\eqref{eq:sgdirect}   is that   if we choose $\th(0) = [c,c,c]$ for any  $c\in \reals$, then all elements of $\th(k)$ are identical, regardless of $\tht$.

There are two takeaways from this  numerical  example.  First, despite the  anonymization, one can still consistently estimate the true parameter set $\tht$.  Second, the   objective~\eqref{eq:multiobj0} is non-convex in $\th$ but convex - so a classical stochastic gradient algorithm can gets tuck in a local minimum. But since the objective is convex in the polynomial coefficients $\lam$, which are constructed as pseudo-observations via the symmetric transform, algorithm~\eqref{eq:lmsbank} converges to the global minimum. 

\subsubsection*{Example 2: Recursive Maximum Likelihood vs Symmetric Transform}

The recursive EM algorithm (\eqref{eq:rem} (REM) in Appendix) requires knowledge of the noise distribution and  probabilities of permutation process~$\state$.  When these are known,   REM performs extremely well. But in the {\em mis-specified case}, where the assumed noise distribution is different to the actual  distribution,   REM can yield a significant bias in the estimates.

\begin{figure*}[p]
  \centering
  \begin{subfigure}{0.4\linewidth}
    \includegraphics[scale=0.4]{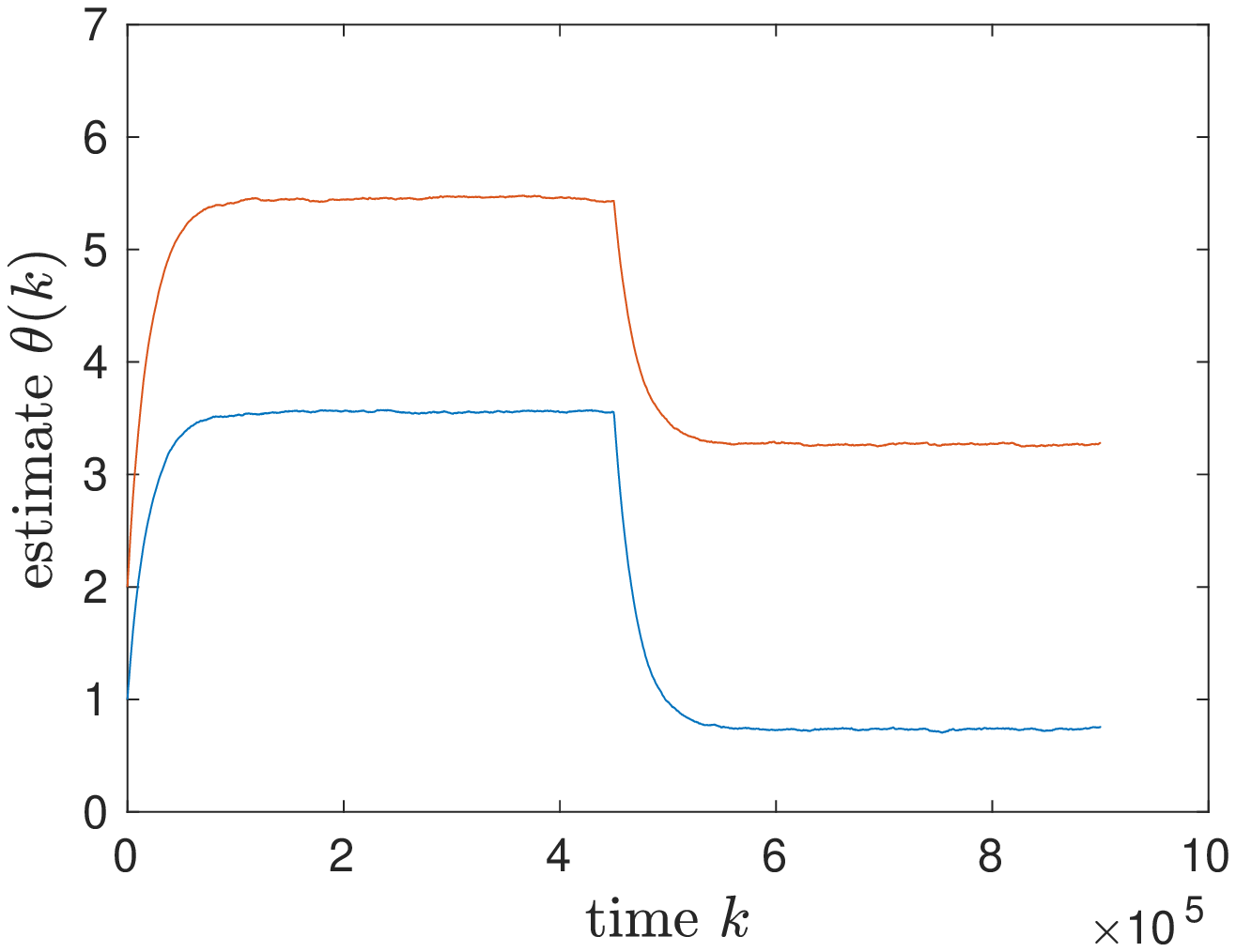}
 \caption{Recursive EM Algorithm~\eqref{eq:rem}.}
    \label{fig:rem}
  \end{subfigure}   \hspace{1cm}
   \begin{subfigure}{0.4\linewidth}
    \includegraphics[scale=0.4]{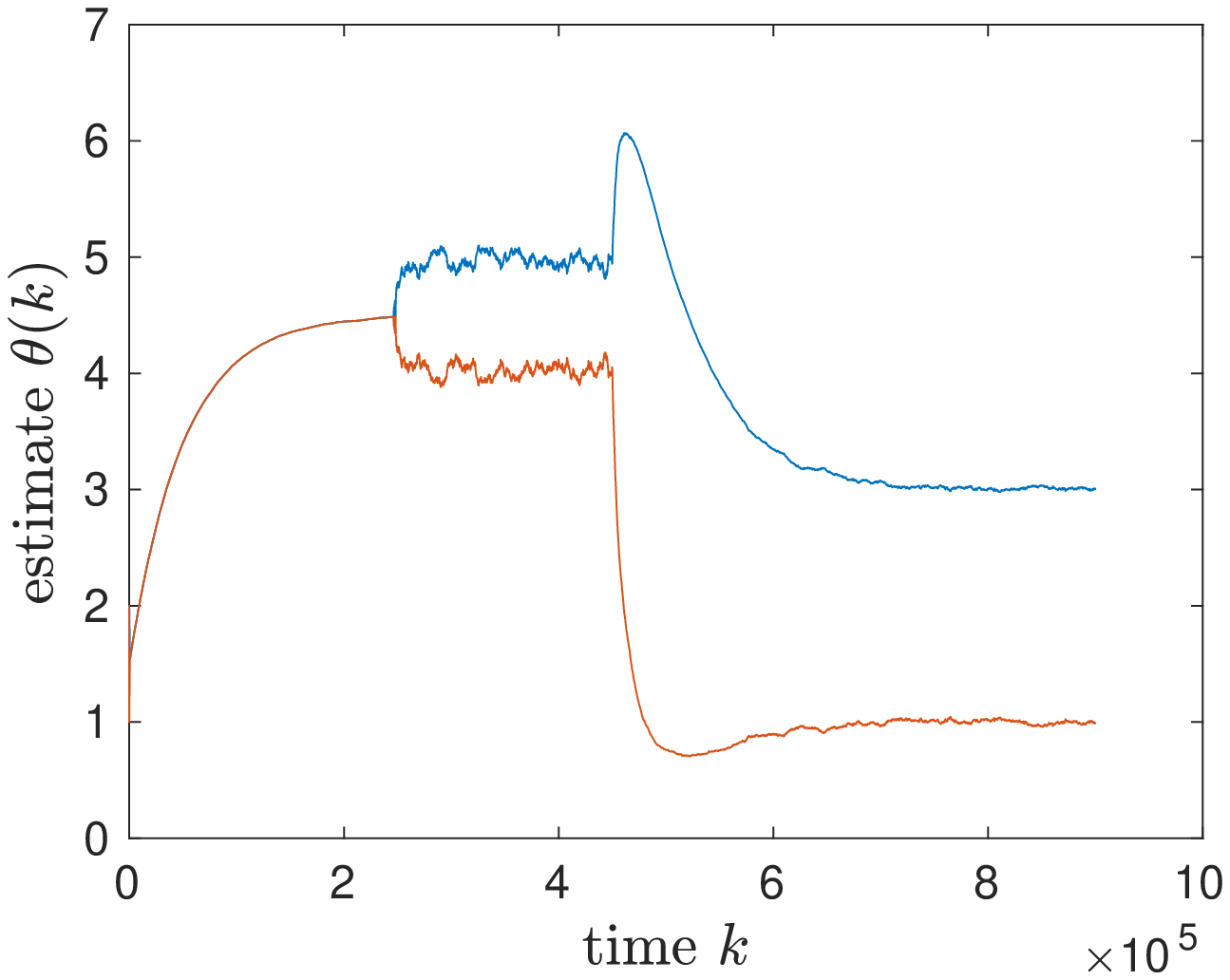}
 \caption{Adaptive Filtering  algorithm~\eqref{eq:lmsbank}.}
    \label{fig:mlevslms}
  \end{subfigure} 
  \caption{Recursive Expectation Maximization algorithm vs Symmetric Transform based Adaptive Filtering algorithm. Both algorithms  operate on anonymized observations~\eqref{eq:obs}, \eqref{eq:setobs} corrupted by Laplacian noise. The true parameter is $\tht=[4,5]^\p$. The  recursive EM shows  a significant bias in the mis-specified case; in comparison the symmetric transform based algorithm converges to the true parameter value but the convergence is slower. The parameters are specified in Example 2.}
\end{figure*}

 We simulated  anonymized observations~\eqref{eq:obs}, \eqref{eq:setobs} for $\thdim=1,\numtarget=2$ with zero mean iid Laplacian noise $\err$ with standard deviation 2.    The true parameter is $\tht = [4,5]$ for $k \leq 3\times 10^5 $ time points  and then changes to
$[1,3]$.   We ran REM~\eqref{eq:rem}  assuming unit variance Gaussian noise.
The step size $\ep = 5 \times 10^{-5}$ and initial estimate $\th(0)=[1,2]^\p$.
Figure~\ref{fig:rem} shows that the algorithm yields a significant bias in the estimate
for $\tht$; the estimates $\th(k)$ converge to $[3.5590,5.4559]^\p$ for the first $3 \times 10^5$ points and then to $[0.7405,3.2658]^\p$.

We then computed the pseudo-observations~\eqref{eq:pseudo} using the scalar symmetric transform~\eqref{eq:lam} and ran the adaptive filtering algorithm~\eqref{eq:lmsbank} with step size $\ep=2\times 10^{-5}$ and initial condition $\th(0)=[1,2]^\p$. Figure~\ref{fig:mlevslms} displays  the sample path estimates $\th(k)$.
We see empirically that the convergence  of adaptive filtering algorithm is slower than the recursive EM, but the estimates converge to the true parameter $\tht$ (with no bias).


\subsubsection*{Example 3: Symmetric Transform for Vector case. $\thdim =2$, $\numtarget =2$}
We consider $\numtarget = 2$ independent processes each of dimension $\thdim = 2$ with anonymized observations generated  by~\eqref{eq:setobs}.
The true models that generate the observations for the two independent processes via~\eqref{eq:obs} are  $\tht_1=[-2,\; 6]^\p$, $\tht_2=[4, \; 5]^\p$.
The $2 \times 2$ input regression matrix in~\eqref{eq:obs} was chosen with iid elements  $\reg_{ij}(k) \sim \normal(0,1)$.
The 2-dimensional noise error vector $v(k)$ has iid elements $ \normal(0,\sigma_v^2)$ where $\sigma_v = 10^{-1}$.

\begin{figure*}[p]
  \centering
   \centering
  \begin{subfigure}{0.4\linewidth}
    \includegraphics[scale=0.4]{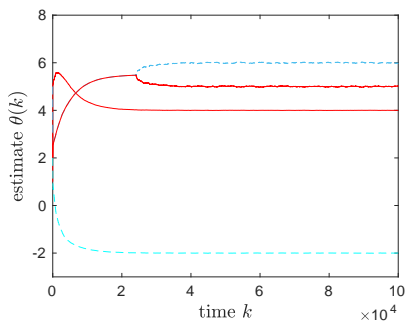}
 \caption{Algorithm~\eqref{eq:lmsvector} operating on vector symmetric transforms converges to global optimum $\tht$.}
    \label{fig:vectlam}
  \end{subfigure} \hspace{1cm}
  \begin{subfigure}{0.4\linewidth}
    \includegraphics[scale=0.4]{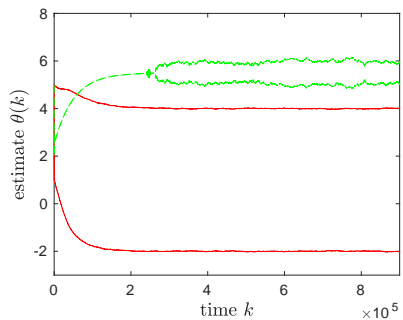}
 \caption{Algorithm~\eqref{eq:lmsvector} operating on pseudo-observations generated by the naive symmetric transform~\eqref{eq:naivesym} converges to the ghost process parameters $\{[-2,\; 5]^\p, [4,\;6]^\p\}$  instead of  the true model set $\{[-2,\; 6]^\p, [4,\;5]^\p\}$.}
    \label{fig:naivelam}
  \end{subfigure} \hspace{1cm}
  \caption{Anonymized estimation problem in Example 3 of Sec.~\ref{sec:numerical}. }
  \label{fig:vectorsim}
\end{figure*}

Given the anonymized observations, we constructed the pseudo-observations using the vector symmetric transform~\eqref{eq:multi}. 
We ran the adaptive filtering  algorithm~\eqref{eq:lmsvector} with step size
$\ep = 10^{-4}$ 
on these pseudo-observations. Figure~\ref{fig:vectlam} shows that the  estimates  converge to the true model set $\tht$.

Next we constructed the naive pseudo observations from the anonymized observations by using the naive transform $\bsym$~\eqref{eq:naivesym}.
We then ran the adaptive filtering  algorithm~\eqref{eq:lmsvector} with step size $\ep=10^{-5}$ on these
naive pseudo-observations.
We see from Figure~\ref{fig:naivelam} that the estimates  converge
to $\{[-2,\; 5]^\p, [4,\;6]^\p\}$  instead of  the true model set $\{[-2,\; 6]^\p, [4,\;5]^\p\}$. So naively  applying the scalar symmetric transform element-wise can result in estimates that swap the elements of $\tht$.
In comparison, the vector symmetric transform together with algorithm~\eqref{eq:lmsvector} yield consistent estimates of $\tht$.

\subsubsection*{Example 4: Mid-sized Example}
In Appendix~\ref{sec:appsim} (supplementary document) 
we consider the case $\numtarget=4$ and $\thdim=10$.
We show that algorithm~\eqref{eq:lmsvector} successfully estimates the   parameters. In comparison,  the naive symmetric transform loses ordering information resulting in ghost process estimates.

\section{Conclusions}

We  proposed  a symmetric transform based adaptive filtering algorithm  for parameter estimation when the observations are a set (unordered) rather than a vector. Such observation sets arise due to uncertainty in sensing or deliberate  anonymization of data.
By exploiting the uniqueness of factorization over polynomial rings, Theorems~\ref{thm:scalar} and~\ref{thm:vector} showed  that the adaptive filtering algorithms  converge to the true parameter (global minimum).
Lemma~\ref{lem:effloss}  characterized the loss in efficiency due to anonymization by evaluating the asymptotic covariance of the algorithm via the algebraic Liapunov equation. Theorem~\ref{thm:hyper} characterized the mean squared error when the underlying true parameter evolves over time according to an unknown Markov chain.  Finally Theorem~\ref{thm:blackwell}  related the asymptotic covariance (convergence rate) of the adaptive filtering algorithm to a Bayesian interpretation of  anonymity of the observations  via mean preserving  Blackwell dominance.

The tools used in this paper, namely symmetric transforms to circumvent data association,  polynomial rings to characterize the attraction points of an adaptive filtering (stochastic gradient) algorithm, and Blackwell dominance to relate a Bayesian interpretation of anonymity to the convergence rate of the adaptive filtering algorithm, can be extended to other formulations.
In future work, it is worth addressing distributed methods for learning with unlabeled data, for example, \cite{VS21} proposes powerful distributed methods.  Also the effect of quantizing  the anonymized data can be studied using~\cite{WY07}.

{\bf Supplementary Document}. The supplementary document contains all proofs,
additional simulation examples and description of a recursive maximum likelihood estimator for
$\tht$. 

\clearpage
\newpage

\appendix

\begin{center} {\large {\bf Supplementary Document}} \\  Adaptive Filtering Algorithms for Set-Valued Observations--Symmetric Measurement Approach to Unlabeled and Anonymized Data \\
  by Vikram Krishnamurthy

\end{center}
\section{Appendix} \label{sec:appendix}

\begin{abstract}
   This supplementary document contains:
   \begin{compactenum}
     \item Description  of the maximum likelihood estimator of the parameter $\tht$ via a recursive Expectation Maximization algorithm in Sec.~\ref{sec:em}.
     \item    Proofs of the theorems stated in the main paper.
\item A medium sized simulation example illustrating the adaptive filtering algorithm and ghost processes in Sec.\ref{sec:appsim}
  \item 
    Example of the symmetric transform $\sym$ for $\thdim=3, \numtarget =3$ in
    Sec.\ref{sec:st3}.
  \end{compactenum}
\end{abstract}

\subsection{Maximum Likelihood Estimation} \label{sec:em}

This section discusses maximum likelihood (ML) estimation of $\tht$ given observations generated by~\eqref{eq:obs}, \eqref{eq:setobs}. The results of this section are not new - they are  used to benchmark the symmetric transform based algorithms derived in the paper.

To give some context, we mentioned in the Introduction that given an observation set $\obs$ (instead of a vector), feeding  it in an arbitrary order into a bank of LMS algorithms will not  converge to $\tht$ in general.
A more sophisticated approach is to order the elements of the observation set at each time based on an estimate of the permutation map $\perm_k$.    
We  can interpret the recursive MLE algorithm below as computing the posterior of  $\perm_k$ and then feeding it into a stochastic gradient algorithm.

Before proceeding it is worthwhile to summarize 
the disadvantages of the MLE approach of this section:   
 \begin{compactenum}
 \item The density function of the noise process $\err$ in ~\eqref{eq:obs} and the probability law of the random process $\state$ in~\eqref{eq:probmodel} need to be known. For example if $\state$ was an iid process, the in principle one can recursively estimate the probabilities of $\state$. However if $\state$ is an arbitrary non-stationary process, then the MLE approach is not useful. 
   
 \item The state space dimension of $\state$ is $\numtarget!$, i.e.,  factorial in the number of processes $\numtarget$.  In comparison, for the symmetric function approach, the number of coefficients of the symmetric transform polynomial is  $O(\numtarget^2)$, see~\eqref{eq:multi}.

 \item The likelihood is not necessarily concave in   $\th$ and so computing the global maximum of the likelihood can be intractable. However, when 
$\err$ in \eqref{eq:obs} is Gaussian, then  \eqref{eq:probmodel}, \eqref{eq:prior},
imply that the   likelihood is concave in $\th$.

\item Why not use the MLE approach together with the symmetric transform? This is not tractable since after applying the symmetric transform, the noise distribution has   complicated form~\eqref{eq:expansionm} that is not amenable to MLE.
\end{compactenum}

We  assume that the permutation process $\state$ in~\eqref{eq:probmodel} is an $\numtarget!$ Markov chain with known transition matrix
 \begin{equation}
   \label{eq:prior}
 P(\state(k+1) = j\, |\, \state(k)=i) = \tp_{ij}  , \quad i,j \in \statespace
\end{equation}
Then~\eqref{eq:probmodel} is a Hidden Markov model (HMM) or dynamic mixture model.
  Notice that  the matrix valued
 observations $\obs(k)$  are generated as  random (Markovian)  permutations of the rows of matrix $\tht \reg(k)$ corrupted by noise. Given these observations, the aim is to estimate the matrix~$\tht$. 

 In this section,  our aim is to compute the MLE  for $\tht$. 
Given $\horizon$ data points,
 the MLE  is defined  as $\mle= \arg\sup_{\th \in \Th} \log p(\obs(1),\ldots\obs(\horizon) ; \th)$.
 We assume that $\Th$ is a compact subset of $\reals^{\numtarget \times \thdim}$ and so the MLE is
 \begin{equation}
   \label{eq:mle}
   \begin{split}
     \mle &= \argmax_{\th \in \Th} \log p_\th(\obs(1\colon\horizon) ), \\
     \text{ where } &\obs(1\colon\horizon) \defn (\obs(1),\ldots,\obs(\horizon))
   \end{split}
 \end{equation}
Under quite  general conditions the MLE $\mle$ of a HMM is strongly consistent  (converges w.p.1 to $\tht$) and efficient (achieves the Cramer-Rao lower bound), see~\citeA{CMR05}. 

{\em Remark}. With suitable abuse of notation,
 note that $\obs(k)$ in~\eqref{eq:probmodel} is a matrix, whereas $\obs(k)$ in~\eqref{eq:setobs} is a set.
In the probabilistic setting that we now consider, this distinction is irrelevant. For example,  we could have denoted the anonymization operation~\eqref{eq:setobs} as choosing amongst the permutation matrices with equal probability $1/\numtarget!$.
 In the symmetric transform formulation in previous sections,
 we did not impose  assumptions on how the elements of the observation set are permuted;  the algorithm~\eqref{eq:lmsvector} was agnostic to the order of the elements in the set $\obs(k)$.
 In comparison, in this section  we postulate that the Markov process $\state$ permutes the observations.

 \subsubsection*{Expectation Maximization (EM) Algorithm}
The process $\state$ is
 the latent (unobserved) data that permutes the observations from the $\numtarget$ processes yielding the matrix $\obs(k)$ in~\eqref{eq:probmodel}.
 The Expectation Maximization (EM) algorithm is a convenient numerical method for computing the MLE when there is latent data. Starting with an initial estimate $\th^0$, the EM algorithm iteratively generates a sequence of estimates $\th^i$, where each iteration $i=1,2,\ldots$ comprises  two steps:\\  
 {\em  Step 1. Expectation step}:
Compute the auxiliary likelihood  
\begin{equation}
  \label{eq:Qdef}
  Q(\th,\th^{i} ) \defn \E\{ \log p_\th(\obs(1 \colon \horizon), x(1 \colon\horizon) | \obs(1\colon\horizon),\th^i\}
\end{equation}
where  $\obs(1\colon \horizon) = (\obs(1),\ldots,\obs(\horizon))$ and $\state(1\colon\horizon)= (\state(1),\ldots,\state(\horizon))$.
In our case, from~\eqref{eq:obs}, \eqref{eq:probmodel}, \eqref{eq:prior}, imply
\begin{equation}
  \label{eq:Q}
  Q(\th,\th^{i} )        = \sum_{k=1}^\horizon \sum_{i=1}^\statedim \belief_i(k|\horizon)\, \log p_\err\big(\obs(k) - \perm(i) \,\reg(k) \,\th \big)
\end{equation}
The smoothed probabilities  $\belief_i(k|\horizon)$ are computed using a forward backward algorithm~\citeA{Kri16}; we omit details here.\\
{\em Step 2. Maximization step}: Compute $\th^{i+1} = \argmax_\th Q(\th,\th^i)$.

Under mild continuity conditions of $Q(\th,\th^i)$ wrt $\th$, it is well known~\citeA{Wu83} that  the EM algorithm climbs the likelihood  surface and converges to a local stationary point $\th^*$ of the log likelihood $\log p(\obs(1)\,\ldots, \obs(\horizon) ;\th)$.

\subsubsection*{Recursive EM Algorithm for Anonymized Observations -- IID Permutations}
We are interested in sequential (on-line) estimation that generates a sequence of estimates $\th(k)$ over time~$k$.  So we formulate a recursive (on-line) EM algorithm. In the numerical examples presented in Sec.~\ref{sec:numerical}, we will consider the case where permuting process $\state$ is iid with $\belief(i) \defn P(\state(k) = i)$, rather than a more general Markov chain.
(Recursive EM algorithms can also be developed for HMMs, but the convergence proof is  more technical.)

Since $\state$ and $ \obs$ are iid processes,
assuming  $| \E_{\tht}\{\E\{ \log p_\th(y(k),x(k))|y(k) ,\bth\}\}| < \infty$, 
it follows from Kolmogorov's  strong law of large numbers that
\begin{equation}
  \label{eq:sllnQ}
  \begin{split}
\lim_{\horizon\rightarrow \infty} \frac{1}{\horizon}  Q(\th,\bth)   =
\lim_{\horizon\rightarrow \infty} \frac{1}{\horizon} \sum_{k=1}^\horizon
\E\{ \log p_\th(y(k),x(k))|y(k) ,\bth\} \\
=
    \E_{\tht}\{\E\{\log p_\th(\obs,\state)|\obs,\bth\}\} \quad \text{ w.p.1 }
  \end{split}
\end{equation}
The recursive EM algorithm is a stochastic gradient ascent algorithm that operates on the above objective:
\begin{equation}
  \label{eq:grem}
\th(k+1) = \th(k) + \ep \, \nabla_\th \E\{\log  p_\th(\obs,\state)|\obs,\th(k)\}\big\vert_{\th = \th(k)}
\end{equation}
where $\ep>0$ is a constant  step size.
Then
starting with initial estimate $\th(0)$, the recursive EM algorithm generates estimates $\th(k), k=1,2\ldots,$ as follows:
\begin{equation}
  \label{eq:rem}
  \begin{split}
    \th(k+1) &= \th(k) + \ep \sum_{i \in \statespace} \belief_i(k) \, \nabla_\th \big[\log p_\err\big(\obs(k) - \perm(i) \, \reg(k)\, \th(k)\big)\big]
    \\
     \belief_i(k) &\propto  \prior(i) \, p_\err \big( \obs(k) - \perm(i) \reg(k) \th(k)\big)
  \end{split}
\end{equation}
So~\eqref{eq:rem} uses a weighted combination  of the posterior probability of all possible permutations to scale the gradient of the auxiliary likelihood $Q$; and these scaled gradients are used in the stochastic gradient ascent algorithm.

{\em Remark}. Let us  explain the rationale behind the recursive EM algorithm~\eqref{eq:rem}. 
First, assuming
$\E_{\tht}|\log p_\th(\obs)| < \infty$, it follows 
by Kolmogorov's strong law of large numbers that  the log likelihood satisfies 
\begin{equation}
  \label{eq:sllnL}
  \lim_{\horizon\rightarrow \infty} \frac{1}{\horizon}\log p_\th(\obs(1\colon\horizon)) = \E_{\tht}\{\log p_\th(\obs) \} \quad \text{ w.p.1 }
\end{equation}
Next  Fisher's identity  relates the gradient of  the log likelihood to that of the auxiliary likelihood $Q$:  
$$ \nabla_\th \log p_{\bth}(\obs(1 \colon \horizon) ) = \nabla_\th  Q(\th,\bth) \big\vert_{\th = \bth}$$
Thus for $\horizon \rightarrow \infty$, it follows from Fisher's identity, \eqref{eq:sllnQ} and~\eqref{eq:sllnL} that
\begin{equation}
\label{eq:fisheravg}
  \nabla_\th \log p_\th(\obs) \big\vert_{\bth} =  \nabla_\th \E\{\log p_\th(\obs,\state)|\obs,\bth\}\}\big\vert_{\th = \bth}
\end{equation}
The  regularity conditions  for~\eqref{eq:fisheravg}  to hold are (i) $L(\th)$ is differentiable on $\Th$. (ii) For any $\bth\in \Th$, $Q(\th,\bth)$ is continuously differentiable on $\Th$. (iii) For any
$\th \in \Th $, both $|\log p_\th(\obs,\state) |\leq \alpha $ and 
$\|\nabla_\th \log p_\th(\obs,\state)\| \leq \beta$ for all $\obs$ with $\E\{\alpha\}< \infty$,
$\E\{\beta\} < \infty$.
Note that (ii) and (iii) are sufficient (via the dominated convergence theorem) for
$ \nabla_\th \E\{\log p_\th(\obs,\state)|\obs,\bth\}\} = \E\{\nabla_\th \log p_\th(\obs,\state)|\obs,\bth\}\} $.

In light of~\eqref{eq:fisheravg}, we see that~\eqref{eq:rem} is  a stochastic gradient algorithm to maximize the objective
$\E_{\tht}\{\log p_\th(\obs) \}$ wrt~$\th$. Moreover, we can rewrite this objective in terms of the Kullback Liebler (KL) divergence:
\begin{equation}
  \label{eq:KLD}
  \begin{split}
   & \argmax_{\th\in \Th} \E_{\tht}\{\log p_\th(\obs) \}  \\ &= \argmin_{\th \in \Th}\E_{\tht}\{\log p_{\tht}(\obs) \} - \E_{\tht}\{\log p_\th(\obs) \}  \\ &= \argmin_{\th\in \Th} D(\tht|| \th)
  \end{split}
\end{equation}
where $D(\tht|| \th)$ is the KL divergence between $p_{\tht}(\obs)$ and $p_\th(\obs)$.
To summarize, the recursive EM algorithm  \eqref{eq:rem} is  a stochastic gradient algorithm to minimize the  KL divergence~\eqref{eq:KLD}.

\vspace{0.5cm}

{\large{\bf Proofs of Theorems}}

\subsection{Proof of Theorem~\ref{thm:scalar}} \label{sec:scalarproof}
  Starting from~\eqref{eq:polydef}, by expanding the symmetric polynomial coefficients we have in polynomial notation  
  \begin{equation}
    \label{eq:expansion}
\begin{split}
    \sym\{\obs_1,\ldots,\obs_\numtarget\}(s) =  \sym\{\reg \tht_1,\ldots, \reg \tht_\numtarget\}(s) +   w(s) , \\  \text{ where } w(s) \defn  \sum_{\I \subseteq \numtargetset, \I \neq \emptyset}
    \prod_{\target \in \I}  \err_\target \,  \sym\{\reg \tht_i, i \in \numtargetset - \I\}(s)
\end{split}
\end{equation}
  The definition of the noise polynomial $w$  in~\eqref{eq:expansion} involves the summation  over all non-empty subsets $\I$ of $\numtargetset$.

  {\em Remark}. To illustrate formula~\eqref{eq:expansion}, consider $\numtarget=2$.
  First expanding out we have
\begin{equation}      \label{eq:ws}
  \begin{split}  
 \sym\{\obs_1,\obs_2\}(s)  &= (s+ \reg \tht_1 + \err_1) (s + \reg \tht_2 + \err_2) \\
& = (s + \reg \tht_1) (s+\reg \tht_2)  + w(s) \\
    w(s) &= s(\err_1 + \err_2) + \reg \tht_1 \err_2 + \reg \tht_2 \err_1 + \err_1 \err_2
  \end{split}
\end{equation}
  Let us verify the expression for  $w(s)$ in~\eqref{eq:expansion}: Since, $\numtargetset = \{1,2\}$,
  for $\I = \{1\}$,  $\sym\{\reg \tht_2\} (s)= (s+ \reg \tht_2)$; for $\I=\{2\}$,
$\sym\{\reg \tht_1\} (s)= (s+\reg\tht_1)$; 
and for $\I = \{1,2\}$,
$\sym\{\emptyset\} = 1$.
Then $w(s) = v_1 \,S\{\reg \tht_2\}(s)+ v_2\, S\{\reg \tht_1\} (s)+ v_1 v_2 \, S\{\emptyset\}$ which yields the above  expression.
Note that for $\I = \emptyset$, $\sym\{\reg \tht_1,\reg \tht_2\}(s) = (s+ \reg \tht_1) (s+\reg \tht_2)$.

From~\eqref{eq:expansion}, since $\err_\target$ are zero mean mutually independent, clearly  $\E\{w(s)\} = 0$; equivalently, the vector $w$ is zero mean, i.e., $\E\{w_\target\} = 0$, $\target\in \numtargetset$.
We can then express~\eqref{eq:expansion} component-wise by reading off the $\numtarget$ coefficients of the polynomial: 
\begin{equation} 
  \begin{split}
    z_\target  &= \sym_\target\{\reg\, \tht_1, \ldots, \reg\, \tht_\target\} + w_\target \\
              &=  \reg^l \, \sym_\target\{\tht_1,\ldots,\tht_\target\} + w_\target  
    = \reg^l \, \lamt_\target + w_\target, \quad l \in \numtargetset
 \end{split} 
\end{equation}
where the second equality follows from~\eqref{eq:scale}.
So we can rewrite objective~\eqref{eq:multiobj0} as
  $\numtarget $ decoupled convex  optimization problems in terms of the variables $\lam$:
\begin{equation} \label{eq:decoupled2}
    \min_{\lam_\target}  \E| z_\target - \reg^l  \lam_l|^2 \quad
    \text{ where }   \quad
      z_l  (k)     = \big(\reg(k)\big)^l\, \lamt_\target + w_\target (k),
      \end{equation}
  Notice that each of the $\numtarget$ objectives  in~\eqref{eq:decoupled2} are quadratic (convex) in $\lam_\target$.
It is easily verified that  $\lam_\target = \lamt, \target \in \numtargetset$ is the unique minimizer that solves~\eqref{eq:decoupled2}.
  Since $\lam = \lamt$ is the unique minimizer of~\eqref{eq:decoupled2} and $\sym$ is uniquely invertible (see~\eqref{eq:ipoly}),  it follows that
$\th = \sym^{-1}(\lam) = \sym^{-1}(\lamt) = \tht$ is the unique minimizer of~\eqref{eq:multiobj0}.

Finally, because each stochastic optimization problem~\eqref{eq:decoupled} is quadratic,  the adaptive filtering algorithm~\eqref{eq:lmsbank} yields estimates $\lam(k)$ that converge  to $\lamt$ in probability. Therefore the unique set of roots $\th(k) = \sym^{-1}(\lam(k))$ converges to $\tht$ in probability.  (Recall from~\eqref{eq:ipoly} that $\th(k)$ are the set of roots of the polynomial with coefficients $\lam(k)$.)

\subsection{Proof of Theorem~\ref{thm:vector}}

{\bf Statement 1}.
Evaluating~\eqref{eq:multi} with $\obs$ in~\eqref{eq:obs} we obtain
  \begin{equation}
    \label{eq:expansionm}
    \begin{split}
 \sym\{\obs_1,\ldots,\obs_\numtarget\}(s,t) =  \sym\{\reg \tht_1,\ldots, \reg \tht_\numtarget\}(s,t) +   w(s,t) \hspace{2cm} \\
      w(s,t) \defn  \sum_{\I \subseteq \numtargetset, \I \neq \emptyset}
      \prod_{\target \in \I}  \big[\sum_{j=1}^\thdim \err_{\target,j}\, t^{j-1} \big]\,  \sym\{\reg \tht_i, i \in \numtargetset - \I\}(s,t)
    \end{split}
  \end{equation}
  The definition of the noise polynomial $w$  in~\eqref{eq:expansionm} involves the summation  over all non-empty subsets $\I$ of $\numtargetset$.

  Since $\err_\target$ are zero mean mutually independent,   it follows from~\eqref{eq:expansionm}
  that  $\E\{w(s,t)\} = 0$; equivalently, the matrix $w$ comprising the coefficients of the polynomial $w(s,t)$   is zero mean, i.e., $\E\{w_{\target,m}\} = 0$, $m \in \M$, $\target\in \numtargetset$.

  {\em Remark.} Since the notation in~\eqref{eq:expansionm} is  complex,  we  illustrate formula~\eqref{eq:expansionm} for the case
  $\numtarget=2$, $\thdim=2$.
Let $\reg_i$ denote the $i$-th row of $\reg$.
  Evaluating~\eqref{eq:multi} with $\obs$ given by~\eqref{eq:obs}, we obtain  the polynomial
  \begin{equation}
    \label{eq:multiexample}
    \begin{split}
      \sym\{\obs\}(s,t) &= (s + \reg_1^\p \tht_1 + t \reg_2^\p \tht_1)\, (s+ \reg_1^\p \tht_2 + t \reg_2^\p  \tht_2) \\ & +w(s,t) \\
    \text{ where }   w(s,t) &= (s + \reg_1^\p \tht_1 + t \reg_2^\p \tht_1)\, (\err_{2,1} + \err_{2,2} t)  \\ & +
    (s+ \reg_1^\p \tht_2 + t \reg_2^\p  \tht_2) \, (\err_{1,1} + \err_{1,2} t)  \\ & + 
      (\err_{1,1} + \err_{1,2} t) \, (\err_{2,1} + \err_{2,2} t)
    \end{split}
  \end{equation}
 We now show that ~\eqref{eq:expansionm} gives the same expression for $w(s,t)$ as~\eqref{eq:multiexample}.
  For  $\numtargetset = \{1,2\}$, we evaluate each subset $\I$ and the corresponding term in~\eqref{eq:expansionm}. For $\I = \{1\}$,  $\sym\{\reg \tht_2\} (s,t)= (s + \reg_1^\p \tht_2 + t \reg_2^\p \tht_2) $ and the multiplying noise term is $\err_{1,1} + \err_{1,2} t$. For $\I = \{2\}$,
  $\sym\{\reg \tht_1\} (s,t)= (s + \reg_1^\p \tht_1 + t \reg_2^\p \tht_1) $ and the multiplying noise term is $\err_{2,1} + \err_{2,2} t$. Finally for $\I = \{1,2\}$ , $\sym\{\emptyset \} = 1$ and the multiplying noise term is $(\err_{1,1} + \err_{1,2} t) \, (\err_{2,1} + \err_{2,2} t)$. Adding these three terms as in~\eqref{eq:expansionm},  we obtain noise polynomial  $w(s,t)$ in~\eqref{eq:multiexample}.
  Note that for $\I = \emptyset$, $\sym\{\reg \tht_1,\reg \tht_2\}(s,t)$, we obtain the 
 signal polynomial $(s + \reg_1^\p \tht_1 + t \reg_2^\p \tht_1)\, (s+ \reg_1^\p \tht_2 + t \reg_2^\p  \tht_2) $.

 {\bf Statement 2}. To keep the notation manageable we prove the result for  $\numtarget =3$,  The general proof is identical but the notation becomes unreadable.

 Suppose $\sym_{\target m}\{\th\} =  \sum_{i, j,k} \th_{1i}\,\th_{2j} \,\th_{3k}$ where this sum is symmetric over indices $ [i,j,k]$ in a certain set. We denote this symmetric sum as $\sym_{lm}\{\th\} = [\th_{1i} \th_{2j} \th_{3k} ] \circ [i,j,k]$. \\
 {\em Example}. If $\thdim=3$, $\numtarget=3$,
 $\sym_{12}(\th) = \th_{11} \th_{21} \th_{32} + \th_{11} \th_{22} \th_{31} + \th_{12} \th_{21} \th_{31} =  [\th_{1i} \, \th_{2j}\, \th_{3k}] \circ[1,1,2]$.

 Then
 $$\sym_{\target m} \{\reg \th\} = \sum_{i, j , k} \reg_i^\p \th_{1}\,\reg_j^\p \th_{2} \,\reg_k^\p \th_{3} = [\reg_i^\p \th_1\, \reg_j^\p \th_2\,\reg_k^\p \th_3]\circ [i,j,k]$$
 We want to show that this yields the RHS of~\eqref{eq:vectorhom}. The trick is to encode the above expression in terms of a polynomial in variable $t$:
 \begin{equation}
   \label{eq:tpoly}
   \begin{split}
 \sym_{\target m} \{\reg \th\}(t) &=
 \sum_{i, j, k} \sum_{p=1}^D \reg_{i,p}^\p \th_{1p} t^{p-1} \,\sum_{q=1}^D \reg_{j,q}^\p \th_{2q} t^{q-1} \\ & \times  \,\sum_{r=1}^D \reg_{k,r}^\p \th_{3r} t^{r-1}  
  \\
     &= [\reg_{ip}\reg_{jq} \reg_{kr} \th_{1p} \th_{2q} \th_{3r} ] \circ [ ([p,q,r]\in t^0) \\ & + ([p,q,r] \in t^1) + ([p,q,r] \in t^2) + \cdots ] \circ [i,j,k]
   \end{split}
 \end{equation}
 where we grouped the symmetric  coefficients of powers of $t$ in the last equation above.
 Clearly $\sym_{\target m}\{\reg \th\} = \sym_{\target m} \{\reg \th\}(1)$, i.e., by setting $t=1$.

 Next examining~\eqref{eq:tpoly}, we see that for each $n$,  the symmetric coefficients of $t^{n-1}$  satisfy   
 \begin{equation}
   \label{eq:regpm}
   \begin{split}
     [ \reg_{ip} \reg_{jq} \reg_{kr} \th_{1p} \th_{2q} \th_{3r}] \circ ([p,q,r] \in t^{n-1}) \circ [i,j,k]  \\ = \sym_{ln}\{\reg^{lm}\}\, \sym_{ln}\{\th\}
   \end{split}
 \end{equation}
$$\reg^{\target m} =
\begin{bmatrix}   \reg_{i,1} & \reg_{j,2} & \reg_{k,3} \\
  \reg_{i,2} & \reg_{j,2} & \reg_{k,3} \\
  \vdots  & \vdots & \vdots \\
  \reg_{i,\thdim}&  \reg_{j,\thdim} & \reg_{k,\thdim} 
\end{bmatrix}
$$
Thus~\eqref{eq:tpoly}, \eqref{eq:regpm}  with coefficients $\lam_{\target n} = \sym_{\target n}(\th)$ yields
\begin{equation*}
  \begin{split}
\sym_{\target m} \{\reg \th\}(t) = \sym_{\target 1}\{\reg^{\target,m}\}\, \lam_{\target 1} +
\sym_{\target 2}\{\reg^{\target,m}\}\, \lam_{\target 2} t  + \ldots  
    \\ +\sym_{\target,\M}\,\{\reg^{\target,m}\} \lam_{\target,\M}  t^{\M}
  \end{split}
\end{equation*}

  {\bf Statement 3}. This follows immediately by substituting~\eqref{eq:multinoise}, \eqref{eq:vectorhom} into~\eqref{eq:multiobj2}.
  
{\bf Statement 4}.  For notational convenience we use 
$\{\reg \th\}$ to denote the set $ \{\reg \,\th_1, \reg\, \th_2,\ldots, \reg \, \th_\numtarget\big\} $.
Let $\co_{\target,m}$, $m \in \M$, $\target \in \numtargetset$ denote the coefficients of the polynomial
$\sym \{\reg \,\th\big\} $.
Using~\eqref{eq:multinoise},  $\co_{\target,m}\{\obs\} = \co_{\target,m}\{\reg\tht + \err\} = \co_{\target,m}\{\reg \tht\} + w_{\target,m}$ where $w_{\target,m}$ is zero mean iid over time; so  we can rewrite
the  estimation objective~\eqref{eq:multiobj2}  as
$$\argmin_\th \E\{ \sum_{\target\in \numtargetset} \sum_{m \in \M}  \| \co_{\target,m}\{\reg \th\} -  \co_{\target,m}\{\reg \tht\} \|^2\}
$$
Clearly the above  minimum is achieved  by choosing  $\th^*$ such that  the polynomial coefficients satisfy  
$$ \co_{\target,m}\{\reg \th^*\} =  \co_{\target,m}\{\reg \tht\}  \;\text{ w.p.1} , \quad  m \in \M,\; l \in \numtargetset$$
Next since $\sym$ is uniquely invertible, applying $\sym^{-1}$ to the polynomial coefficients yields the unique set of roots
$$\{ \reg \th^*_\target, m \in \M,\; l \in \numtargetset\}  = \{\reg \tht_\target  ,  m \in \M,\; l \in \numtargetset\}\;  \text{ w.p.1} $$
for all random variable realizations $\reg$. This in turn implies
$ \{ \th^*_\target, m \in \M,\; l \in \numtargetset\} = \{\tht_\target,m \in \M, l \in \numtargetset\}$.

\textit{Remark.} The reader may wonder why the above proof breaks down for the naive symmetric transform~\eqref{eq:naivesym}. We note that for the naive vector symmetric transform~\eqref{eq:naivesym}, the above proof of Statement 4 does not hold.  Even though 
$\bsym_{\target,j}\{\reg \th^*\} =  \bsym_{\target,j}\{\reg \tht\} $  for all
$\target, j$, it does not follow  that
$\{ \reg \th^*_\target, m \in \M,\; l \in \numtargetset\}  = \{\reg \tht_\target  ,  m \in \M,\; l \in \numtargetset\}$  w.p.1. This is because as discussed  below~\eqref{eq:naiveth}, the naive symmetric transform does not preserve the ordering of  vectors.

\subsection{Sensitivity of Symmetric Transform Polynomial}
Here we derive the expression in~\eqref{eq:senspol}.
\begin{theorem}
\label{thm:sensitivity}
  Suppose $\Bth=\{\th_1,\ldots,\th_L\}$ is the set of factors of the polynomial 
  $\sym\{\Bth\}(s) = s^\numtarget+\sum_{l=1}^\numtarget \lam_l s^{\numtarget-l} $.
  That is,   $\sym\{\Bth\}(s) =
  \prod_{l=1}^L(s+\th_l)$.  Assume $\th\in \Bth$ is a distinct (non-repeated) factor. Then
\begin{equation}
\label{eq:theoremsensitivity}
  \frac{d\th}{d\lam_{l}} =  (-1)^{l+1}\, \th^{L-l}\,
   \bigg[ \frac{d\sym\{\Bth\}(-\th)}{d\th}\bigg]^{-1}
 \end{equation}
\end{theorem}

{\bf Proof}:
The proof follows from the following two lemmas.

\begin{lemma}\label{lem:Spolyroot}
  Suppose $\Bth=\{\th_1,\ldots,\th_L\}$ is the set of factors of the polynomial 
  $\sym\{\Bth\}(s) = s^\numtarget+\sum_{l=1}^\numtarget \lam_l s^{\numtarget-l} $.
  That is,   $\sym\{\Bth\}(s) =
\prod_{l=1}^L(s+\th_l)$. 
Then $\Bth$ is also the set of roots of the polynomial $\sym\{\Bth\}(-s) = s^\numtarget + \sum_{l=1}^\numtarget (-1)^l \lam_l s^{\numtarget-l}$. That is,  for any $\th \in \Bth$, it follows that $\sym\{\Bth\}(-\th)= 0$.
\end{lemma}

\begin{lemma} \label{lem:polyroots}
Suppose $P_\beta(\th) \defn \th^\numtarget + \sum_{m=0}^{\numtarget-1} \beta_m \,\th^{m} = 0$, i.e., $\th$ is a root of the polynomial $P_\beta(\th)$. Assume $\th$ is a distinct (non-repeated) root. Then
\begin{equation}
\label{eq:betapoly}
\frac{d\th}{d\beta_m} = -\th^m\,  \bigg[ \frac{dP_\beta}{d\th}\bigg]^{-1}
\end{equation}
\end{lemma}

We now  use Lemma~\ref{lem:polyroots} with Lemma~\ref{lem:Spolyroot} to obtain an expression for $d\th/d\lam_l$.  Note that  $\sym\{\Bth\}(-\th) = P_\beta(\th)$  by choosing $\beta_m =(-1)^{\numtarget-m} \lam_{\numtarget-m}$ for $m=0.\ldots, L-1$.
%
Then from Lemma~\ref{lem:polyroots},
\begin{equation}
\label{eq:derivlam}
\frac{d\th}{d\lam_{L-m}} = - (-1)^{L-m}\, \th^m\,  \bigg[ \frac{dP_\beta}{d\th}\bigg]^{-1}
\end{equation}
\begin{equation*} 
  \begin{split}  
\frac{dP_\beta}{d\th} &=(L-1) \th^{L-1} + \sum_{m=0}^{L-1} m \,\beta_m \,
\th^{m-1}
\\ &= (L-1) \th^{L-1} + \sum_{m=0}^{L-1} m (-1)^{L-m} \, \lam_{L-m} \,\th^{m-1}  \text{(Lemma~\ref{lem:Spolyroot})}\\
    &= (L-1) \th^{L-1} + \sum_{l=1}^L (L-l)\,  (-1)^l\, \lam_l \, \th^{L-l-1}  \\ 
\intertext{choosing $  l = L-m $}  &=   \frac{d\sym\{\Bth\}(-\th)}{d\th}  
  \end{split}
\end{equation*}
Therefore plugging $l = L-m$ in~\eqref{eq:derivlam}, we obtain~\eqref{eq:theoremsensitivity}.

\subsection{Proof of Lemma~\ref{lem:map}}
The MAP estimate is correct when the event
$I(\hat{\state}(k) = \state(k))$ occurs. Denoting $i^* = \argmax_i \belief_i$,
the conditional probability that the MAP estimate is correct  given observation $\obs(k)$ is
$$ \E\{ I(\state(k) = i^*)| \obs(k)\} = \sum_{i=1}^\statedim \belief_i(k) \, I(i^* = i) = \max_i \belief_i(k)$$
The error event is
$1 - I(\state(k) = \hat{\state}(k))$.  Therefore the error probability of the MAP estimate is
$  1 - \max_i \belief_i(k)= 1 - \max_i e_i^\p\,\filter(\belief,\obs(k))$.
Finally,   the expected error probability of the MAP estimate  over all possible realizations of $\obs$ is
\begin{equation}
  \label{eq:avgerror}
  \Pavg(\belief;\oprob)  = \inty \big(1 - \max_i e_i^\p\filter(\belief,\obs) \big)\, \filterd(\belief,\obs) =
  1 -\inty   \max_i e_i^\p\oprob_{\obs}\, \belief  
\end{equation}
Here  $\inty$ denotes integration wrt Lebesgue measure when $\obs \in \reals^{\numtarget\times\thdim}$ or counting measure when $\obs$ is a subset of integers.

\subsection{Proof of Theorem~\ref{thm:blackwell}}
{\bf Statement~1.}  
  Since $\obs_\target$ are independent wrt $\target$,  $\cov_\oprob(\obs) \preceq \cov_{\boprob}(\obs)$ implies $\var_\oprob(\obs_\target) \leq \var_{\boprob}(\obs_\target)$. Next
  $\cov(\sym\{\obs\}) $ is a $\numtarget\times \numtarget$ diagonal matrix with $\target$ element
  $\sum_{i_1< i_2<\cdots < i_\target} \var{\obs_{i_1}}\,\var{\obs_{i_2} }\cdots \var{\obs_{i_\target}} $. This together with $\var_\oprob(\obs_\target) \leq \var_{\boprob}(\obs_\target)$ implies
  $\cov_\oprob(\sym\{\obs\})
  \preceq \cov_{\boprob}(\sym\{\obs\})$.

 \textbf{Statement 2.} The proof below exploits that facts that  $\Pavg(\belief,\oprob) = 1 - \sum_\obs \max_i \oprob_\obs \belief$ is concave in $\belief$, and that \ref{a:blackwell}, namely,  $\oprob \blackwell \boprob$ holds.
In particular, \ref{a:blackwell} implies the following factorization of Bayes formula:
  $$ \filter(\belief,\bobs;\boprob) = \inty \filter(\belief,\obs; \oprob)\,
\frac{  \filterd(\belief,\obs;\oprob)}{\filterd(\belief,\bobs;\boprob)}\, \kerM_{\obs,\bobs}$$
where $\filterd(\belief,\bobs;\boprob) = \sum_\obs \filterd(\belief,\obs;\oprob)\, \kerM_{\obs,\bobs}$.
So $ \frac{  \filterd(\belief,\obs;\oprob)}{\filterd(\belief,\bobs;\boprob)}\, \kerM_{\obs,\bobs}$ qualifies as a measure wrt $\obs$.
Since $\Pe(\filter(\belief,\bobs,\boprob)) = 1 - \max_i \filter(\belief,\bobs;\boprob)$  is concave in $\belief$, it follows using Jensen's inequality that
\begin{equation*} 
  \begin{split}  
\Pe(\filter(\belief,\bobs;\boprob)) &= \Pe(\inty \filter(\belief,\obs; \oprob)\,
                                      \frac{  \filterd(\belief,\obs;\oprob)}{\filterd(\belief,\bobs;\boprob)}\, \kerM_{\obs,\bobs}) \\
    &\geq \inty  \Pe( \filter(\belief,\obs; \oprob))\,  \frac{  \filterd(\belief,\obs;\oprob)}{\filterd(\belief,\bobs;\boprob)}\, \kerM_{\obs,\bobs}
\end{split} 
\end{equation*}
So cross multiplying by $\filterd(\belief,\bobs;\boprob)$ and  integrating wrt 
$\obs$ implies  \\
$$\inty \Pe(\filter(\belief,\bobs;\boprob)) \, \filterd(\belief,\bobs;\boprob) 
\geq \inty \Pe(\filter(\belief,\obs;\oprob)) \, \filterd(\belief,\obs;\oprob)$$
which in turn implies
$  \Pavg(\belief;\boprob) \geq \Pavg(\belief;\oprob) $.

  {\bf Statement 3.} 
 Blackwell's classic paper~\citeA[Theorem 3]{Bla53} shows that~\ref{a:blackwell} and
  \ref{a:zeromean}  imply that 
  $\sum_{\obs_{\target m}} \oprob_{i\obs_{\target m}} \, \obs_{\target m}^2   \leq \sum_{\obs_{\target m}}  \boprob_{i\obs_{\target m}} \, \obs_{\target m}^2 $, i.e.,
  $\cov_\oprob(\obs) \leq \cov_{\boprob}(\obs)$. Here we give the proof in more transparent notation. Below we omit the $l,m$ subscripts. Using Blackwell dominance~\ref{a:blackwell}, it follows  that
$$ \inty \obs \boprob_{i y}  = \inty \obs \intyb \oprob_{i,\bobs}\, \kerM_{\bobs,\obs} = \intyb \oprob_{i,\bobs} \inty  \obs \kerM_{\bobs,\obs}
$$
by Fubini's theorem assuming $\inty |\obs| \boprob_{iy} < \infty$.
The mean preserving assumption~\ref{a:zeromean}  implies that the above expression equals
$\intyb \bobs \oprob_{i\bobs} $. Therefore Blackwell dominance~\ref{a:blackwell} and mean preserving spread~\ref{a:zeromean}  imply that the kernel $\kerM$ satisfies
  \begin{equation}
    \label{eq:kerm}
    \inty \obs \kerM_{\bobs,\obs} = \bobs
  \end{equation}
  Next for  any convex function $\fun$, applying Blackwell dominance, it follows that
\begin{equation} 
  \begin{split}  
  \inty \fun(\obs) \,\boprob_{i,\obs} &= \intyb \oprob_{i \bobs} \inty  \fun(\obs) \kerM_{\bobs,\obs} \\
                                          &  \geq \intyb \oprob_{i \bobs} \, \fun(\inty \obs \,\kerM_{\bobs,\obs} )
                                             \quad \text{ (by Jensen's inequality)}
\\ &= 
  \intyb \oprob_{i \bobs} \, \fun(\bobs)  \quad \text{ (by~\eqref{eq:kerm})}
  \end{split}
\end{equation}
So choosing $\fun(\obs) = \obs^2$, it follows that  $\cov_\oprob(\obs) \leq \cov_{\boprob}(\obs)$.

  Therefore, from Statement 1, we have,
  $\cov_\oprob(\sym\{\obs\}) \leq \cov_{\boprob}(\sym\{\obs\})$, or equivalently,
   $R(\oprob) \preceq R(\boprob)$ (wrt positive definite ordering).
  
 Next, it can be shown \citeA{AM79} by  differentiation that the solution of the algebraic Lyapunov equation~\eqref{eq:Lyapunov} satisfies    
  $$ \kalmancov(\oprob) = \int_0^\infty \exp\big(s \Q \big) \, R(\oprob)\, \exp\big(s \Q  \big) \, ds
  $$
  So clearly if $R(\oprob) \preceq R(\boprob)$, then $\kalmancov(\oprob) \preceq \kalmancov(\boprob) $.
  Finally, since $\bSig = {\nabla \sym^{-1}}^\p \Sig \nabla \sym^{-1}$, it follows that
  $\bSig(\oprob) \preceq \bSig(\boprob)$.

  \subsection{Simulation Example. Noisy Matrix Permutation}
\label{sec:appsim}
The aim of this section is to provide a medium-sized numerical example of estimating
$\tht$ with the vector symmetric transform and adaptive filtering algorithm~\eqref{eq:lmsvector}. We also show that applying the naive symmetric transform~\eqref{eq:naivesym} element wise (as opposed to the vector symmetric transform)   loses order information.

  We consider the case $\numtarget=4$ and $\thdim=10$.
The true parameter is 
$$
\tht =\begin{bmatrix}
        1 &      3 &      4 &      5 &      7 &      9 &     10 &     11 &     12 &     13 \cr
        2 &      4 &      5 &     10 &      8 &      7 &      1 &      8 &      9 &     10 \cr
        3 &      1 &      2 &      7 &      6 &      5 &      4 &      5 &      7 &      9 \cr
        6 &     12 &     18 &     24 &     36 &     43 &     50 &     10 &      1 &      3
  \end{bmatrix}
  $$
  The regression matrix $\reg(k)$ was chosen as $I_{3\times 3}$.  The anonymized $\thdim$-dimension observation vectors  were generated according 
  to~\eqref{eq:obs}, \eqref{eq:setobs}.
  
The pseudo observation vectors are constructed at each time $k$ using~\eqref{eq:convolution} as
\begin{equation}
  \label{eq:convexample}
  \begin{split}
  z_1 &= \obs_1 + \obs_2 + \obs_3 + \obs_4, \\
  z_2 &= \obs_1\conv \obs_2 + \obs_1\conv\obs_3 + \obs_1 \conv \obs_4 + \obs_2 \conv \obs_3 + \obs_2 \cov \obs_4 \\ & + \obs_3 \conv \obs_4, \\
  z_3 &= \obs_1 \conv \obs_2 \conv \obs_3 + \obs_1 \conv \obs_2 \conv \obs_4 + \obs_1 \conv \obs_3 \conv \obs_4 \\ & + \obs_2 \conv \obs_3 \conv  \obs_4, \\
 z_4 &= \obs_1 \conv \obs_2 \conv  \obs_3 \conv \obs_4
  \end{split}
\end{equation}
where $\conv$ denotes the  convolution operator and each $\obs_i(k) \in \reals^\thdim$, $i=1,\ldots,4$.

We ran 100 independent trials of the 
 adaptive filtering algorithm~\eqref{eq:lmsvector} on 100 independent pseudo observation sequences.
We computed the relative error of the average estimate $\th^{\text{avg}}(k)$ over the 100 trials  at time $k=50,000$:
$$|\th_{ij}^{\text{avg}}(k) - \tht_{ij}|/\tht_{ij} \leq 7 \times 10^{-4} $$
Thus algorithm~\eqref{eq:lmsvector},  based on the vector symmetric transform, successfully estimates the   parameters.

Next, we ran adaptive filtering algorithm using the naive symmetric transform~\eqref{eq:naivesym}. We see from the  estimate $\th(k)$ at $k=50.000$ below, that all order information is lost
(the boxes indicate the nearest  estimates to the first row of $\tht$):
  \begin{multline*}
  \begin{bmatrix}
   \fbox{1.0052} & 1.0053 & 1.9923 & \fbox{5.0129} & 6.0105 & 5.0033  \cr
   2.0041 & \fbox{2.9971} & \fbox{4.0048} & 7.0028 & \fbox{6.9913} & 7.0086  \cr
   3.0023 & 4.0016 & 4.9988 & 9.9934 & 8.0095 & \fbox{9.0024}  \cr
   5.9997 & 12.0000 & 17.9965 & 24.0002 & 36.0066 & 43.0028  
  \end{bmatrix}
\\
\left.   \begin{matrix}
 0.9939 & 4.9955 & 1.0032 & 2.9986 \\
 3.9995 & 7.9970 & 6.9999 & 8.9985 \\
 \fbox{9.9936} & 9.9942 & 8.9969 & 10.0001 \\
 50.0024 & \fbox{10.9959} & \fbox{12.0031} & \fbox{12.9960} 
  \end{matrix} \right]
\end{multline*}

We found in numerical examples that when rows of $\tht$ are different from each other, the naive transform is able to estimate the order; but when the elements of two rows are close, then the estimate can switch rows resulting in a ghost estimate.

\subsection{Symmetric Transform for $\thdim=3,\numtarget=3$.}
\label{sec:st3}
This final section of the supplementary document gives a complete evaluation of the
 symmetric transform $\sym$ for $\thdim=3, \numtarget=3$ to
 illustrative~\eqref{eq:convolution}  in the paper.
 
Recall $\lam_{\target,m}$ is the coefficient of $s^{\target-1} t^{m-1}$ in the polynomial  $\sym\{\th\}(s,t)$:

\begin{tabular}{L|  L}
\lam_{11} =\sigblank[1,1,1]  & \lam_{21} = \sigblank[0,1,1] \\
\lam_{12} =\sigblank[1,1,2] &  \lam_{22} = \sigblank[0,1,2] \\
  \lam_{13} = \sigblank[1,1,3]+\sigblank[1,2,2] & \lam_{23} = \sigblank[0,1,3]+\sigblank[0,2,2] \\
  \lam_{14} = \sigblank[1,2,3]+\sigblank[2,2,2] & \lam_{24} = \sigblank[0,2,3]   \\
  \lam_{15} = \sigblank[1,3,3]+\sigblank[2,2,3] & \lam_{25} = \sigblank[0,3,3]  \\
  \lam_{16} = \sigblank[2,3,3] &   \\
  \lam_{17} = \sigblank[3,3,3] &  
\end{tabular}
$ \lam_{31} = \sigblank[0,0,1] $, $\lam_{32} = \sigblank[0,0,2] $,  $\lam_{33} = \sigblank[0,0,3]$.

To explain the  compact notation above:
$[a,b,c] = \sum_{\perm\{a,b,c\}} \th_{1a} \th_{2b} \th_{3c} = \th_{1a} \th_{2b} \th_{3c} + \th_{1a} \th_{2c} \th_{3b} + \th_{1b} \th_{2a}\th_{3c} + \th_{1b} \th_{2c} \th_{3a} + \th_{1c} \th_{2a} \th_{3b} + \th_{1c} \th_{2b} \th_{3a} $ 
and we set $\th_{i0} = 1$ for all $i$.

  So 
$\lam_{11} = \sigblank[1,1,1] = \th_{11} \, \th_{21} \, \th_{31}$ since $\perm\{1,1,1\} = \{1,1,1\}$. Also,
 $\lam_{12}=\sigblank[1,1,2] $ is constructed by taking  all permutations of $[1,1,2]$; so 
$\lam_{12} = \th_{11} \th_{21} \th_{32} + \th_{11} \th_{22} \th_{31} + \th_{12} \th_{21} \th_{31} $.
Similarly,  $\lam_{23} = \sigblank[0,1,3]+\sigblank[0,2,2]= 
\th_{11} \th_{23}+ \th_{21} \th_{33} + \th_{11} \th_{33} + \th_{12} \th_{22} + \th_{12} \th_{32} + \th_{22} \th_{32}$ since $\th_{i0} = 1$ by convention.

In the convolution notation of~\eqref{eq:convolution} we can express
$\lam_1 \in \reals^7, \lam_2 \in \reals^5, \lam_3 \in \reals^3$ 
as:
$$\lam_1 = \th_1 \conv \th_2 \conv \th_3,
\;
\lam_2 = \th_1 \conv \th_2 + \th_1 \conv \th_3 + \th_2 \conv \th_3, \;
\lam_3 = \th_1 + \th_2 + \th_3
$$


\end{document}